\documentclass[11pt,a4paper]{article}

\pdfoutput=1
\usepackage{jcappub}
\usepackage{graphicx}
\usepackage{amssymb}
\usepackage{amsmath}
\usepackage{mathtools}
\usepackage{bm}
\usepackage{latexsym}
\usepackage{epsfig}
\usepackage{float}
\usepackage{textcomp}
\usepackage{color}
\usepackage{float}
\usepackage{blindtext}
\usepackage{enumitem}
\usepackage{xcolor}
\usepackage[section]{placeins}
\usepackage{caption}
\usepackage{float}
\usepackage{tikz} % To generate the plot from csv
\usepackage{pgfplots}
\usepackage{subcaption}
\usepackage[labelformat=parens,labelsep=quad,skip=3pt]{caption}
\usepackage{graphicx}
\pgfplotsset{compat=newest} % Allows to place the legend below plot
\usepgfplotslibrary{units} % Allows to enter the units nicely

\usepackage{calligra}
\DeclareMathAlphabet{\mathcalligra}{T1}{calligra}{m}{n}
\DeclareFontShape{T1}{calligra}{m}{n}{<->s*[2.2]callig15}{}

\usepackage[mathscr]{euscript}
\usepackage[export]{adjustbox}
\usepackage{amsmath}
\makeatletter
\gdef\@fpheader{}
\makeatother

\def\Mpl{M_{_{\rm Pl}}}
\def\be{\begin{equation}}
\def\ee{\end{equation}} 
\def\bea{\begin{eqnarray}}
\def\eea{\end{eqnarray}}

\def\lsim{\mathrel{\rlap{\lower4pt\hbox{\hskip0.5pt$\sim$}}
 \raise1pt\hbox{$<$}}}         %less than or approx. symbol
\def\gsim{\mathrel{\rlap{\lower4pt\hbox{\hskip0.5pt$\sim$}}
 \raise1pt\hbox{$>$}}}         %greater than or approx. symbol

\newcommand{\dd}{\mathrm{d}}

\newcommand{\ba}{\begin{aligned}}
\newcommand{\ea}{\end{aligned}}

\newcommand{\x}{\mathbf{x}}

\newcommand{\MPBH}{M_{\rm PBH}}
\newcommand{\Mt}{M(t)}

\newcommand{\MPL}{M_{\rm Pl}}

\def\Mpl{M_{_{\rm Pl}}}
\def\be{\begin{equation}}
\def\ee{\end{equation}} 
\def\bea{\begin{eqnarray}}
\def\eea{\end{eqnarray}}

\definecolor{lime}{HTML}{A6CE39}
\DeclareRobustCommand{\orcidicon}{
	\begin{tikzpicture}
	\draw[lime, fill=lime] (0,0) 
	circle [radius=0.2] 
	node[white] {{\fontfamily{qag}\selectfont \tiny ID}};
	\draw[white, fill=white] (-0.0625,0.095) 
	circle [radius=0.007];
	\end{tikzpicture}
	\hspace{-2mm}
}

\foreach \x in {A, ..., Z}{\expandafter\xdef\csname orcid\x\endcsname{\noexpand\href{https://orcid.org/\csname orcidauthor\x\endcsname}
			{\noexpand\orcidicon}}
}

\newcommand{\ag}[1]{\textcolor{blue}{[Anish: #1]}}

\def\Mpl{M_{\rm Pl}}
\def\be{\begin{equation}}
\def\ee{\end{equation}} 
\def\bea{\begin{eqnarray}}
\def\eea{\end{eqnarray}}

\numberwithin{equation}{section}

\title{
{Distinct signatures of spinning PBH domination and evaporation:\\ {\it doubly peaked gravitational waves, dark relics and CMB complementarity}}\\
}
\author[\ast]{Nilanjandev Bhaumik\orcidA{}}
\author[\dagger]{, Anish Ghoshal\orcidB{}}
\author[\ast]{, Rajeev Kumar Jain\orcidC{}}
\author[\dagger]{and Marek Lewicki\orcidD{}}
\affiliation[\ast]{Department of Physics, Indian Institute of Science, Bangalore 560012, India}
\affiliation[\dagger]{Institute of Theoretical Physics, Faculty of Physics, University of Warsaw,  \\ ul. Pasteura 5, 02-093 ~Warsaw, Poland}
\emailAdd{nilanjandev@iisc.ac.in}
\emailAdd{anish.ghoshal@fuw.edu.pl}
\emailAdd{rkjain@iisc.ac.in}
\emailAdd{marek.lewicki@fuw.edu.pl}

\abstract{
Ultra-low mass primordial black holes (PBH), which may briefly dominate the energy density of the universe but completely evaporate before the big bang nucleosynthesis (BBN), can lead to interesting  observable signatures. In our previous work, we studied the generation of a doubly peaked spectrum of induced stochastic gravitational wave background (ISGWB) for such a scenario and explored the possibility of probing a class of baryogenesis models wherein the emission of massive unstable particles from the PBH evaporation and their subsequent decay contributes to the matter-antimatter asymmetry.
In this work, we extend the scope of our earlier work by including spinning PBHs and consider the emission of light relativistic dark sector particles, which contribute to the dark radiation (DR) and massive stable dark sector particles, thereby accounting for the dark matter (DM) component of the universe. The ISGWB can probe the non-thermal production of these heavy DM particles, which cannot be accessible in laboratory searches. For the case of DR, we find a novel complementarity between the measurements of $\Delta N_{\rm eff}$ from these emitted particles and the ISGWB from PBH domination.
Our results indicate that the ISGWB has a weak dependence on the initial PBH spin. However, for gravitons as the DR particles, the initial PBH spin plays a significant role, and only above a critical value of the initial spin parameter $a_*$, which depends only on initial PBH mass, the graviton emission can be probed in the CMB-HD experiment. Upcoming CMB experiments such as CMB-HD and CMB-Bharat, together with future GW detectors like LISA and ET, open up an exciting possibility of constraining the PBHs parameter space providing deeper insights into the expansion history of the universe  between the end of inflation and BBN.
}

\arxivnumber{}
\begin{document}
\maketitle

%\tableofcontents

%%######################################################################################%%

\section{Introduction}
\label{sec:intro}

The idea of primordial black holes (PBH) was first put forward by Zel’dovich and Novikov \cite{1966AZh....43..758Z}, and later studied in depth by Hawking and Carr  \cite{Hawking:1971ei, Carr:1974nx, Carr:1975qj}. The formation of PBHs in the very early universe has different interesting implications for both cosmology and particle physics \cite{Chapline:1975ojl, Carr:2019kxo, Escriva:2022duf}. 
%Understanding the nature of dark matter (DM) has been one of the cornerstone problems in modern cosmology. 
On the one hand, in the present cosmological landscape of various dark matter (DM) candidates, PBHs have emerged as an important candidate for the cold DM due to the possibility of testing their signatures in gravitational wave (GW)  detectors \cite{Bird:2016dcv,LIGOScientific:2016aoc,LIGOScientific:2020iuh, Abbott:2016blz,Abbott:2016nmj,Abbott:2017vtc, Clesse:2016ajp}. Though the abundance of PBHs is strongly constrained in different mass ranges~\cite{Hutsi:2020sol, Carr:2020gox, Carr:2016hva, Escriva:2022duf, Barnacka:2012bm, Laha:2019ssq, Niikura:2017zjd,EROS-2:2006ryy,Niikura:2019kqi, Ricotti:2007au, Aloni:2016kuh,Poulin:2017bwe, Saha:2021pqf,Mittal:2021egv,Kohri:2022wzp,Hasinger:2020ptw,Tashiro:2012qe,Hektor:2018qqw}, there remains an open window in the asteroid mass range ($10^{-16}-10^{-14} M_{\odot}$) wherein PBHs can be the entire DM~\cite{Montero-Camacho:2019jte}. 
On the other hand, the possibility of PBH formation in the post inflationary universe due to the amplified inflationary scalar perturbations
%The most popular mechanism for PBH formation involves the collapse of large inhomogeneities in the early universe~\cite{ZelNov:1967, Hawking:1971ei, Carr:1974nx, Hawking:1974sw}. However, primordial inflation dilutes all relics, so a relevant abundance of PBHs can form only after inflation.
(for example, from ultra-slow-roll models of inflation \cite{Bhaumik:2019tvl, Garcia-Bellido:2017mdw, Hertzberg:2017dkh, Ballesteros:2017fsr, Ragavendra:2020sop, Mishra:2019pzq}, warm inflation \cite{Arya:2019wck, Bastero-Gil:2021fac, Arya:2023pod}, Gauss-Bonnet theories of inflation \cite{Kawai:2021edk}, two-field models of inflation \cite{Braglia:2020eai, Anguelova:2020nzl, Clesse:2015wea} etc.) has opened up an interesting avenue to constrain the inflationary perturbations modes leaving horizon during last forty efolds of inflation. In such a scenario, PBHs can help in probing the cosmic history between the end of inflation to the start of the Big Bang Nucleosynthesis (BBN), which is not accessible by any other direct observational probes. PBHs have also been suggested to form from strong first-order phase transitions in early universe \cite{Hawking:1982ga, Kodama:1982sf, Jedamzik:1999am, Lewicki:2019gmv}, the collapse of topological defects~\cite{Hawking:1987bn, Polnarev:1988dh, MacGibbon:1997pu, Rubin:2000dq, Rubin:2001yw, Brandenberger:2021zvn}, scalar condensates and topological and non-topological solitons ~\cite{Cotner:2016cvr}, resonant reheating~\cite{Suyama:2004mz}, preheating~\cite{Suyama:2006sr, Bassett:2000ha, Martin:2019nuw,Martin:2020fgl} and confinement of quarks \cite{Dvali:2021byy}. Thus PBHs can help probe or potentially constrain these scenarios. Here we focus on various signatures of ultra-low mass PBHs $(M_{\rm PBH} \lesssim 10^{9} \, {\rm g})$, and how future GW and CMB probes can constrain PBHs parameter space as well as cosmological history before BBN, through these signatures.

Smaller mass PBHs evaporate much faster. As a result, PBHs with mass $M_{\rm PBH} \lesssim 10^{-18}M_{\odot}$ $(M_{\rm PBH} \lesssim 10^{15} \, {\rm g})$ would be completely evaporated by today. BBN provides stringent constraints on the abundance of PBHs in the early universe for masses $10^9\,\textrm{g}\,\leq M_{\textrm{PBH}}\leq 10^{14}\,\textrm{g}$~\cite{Carr:2009jm, Carr:2020gox, Keith:2020jww}, as PBHs of this mass range evaporates during or after BBN. On the other hand, PBHs with mass below $10^9\,\text{g}$ evaporate far before BBN. Therefore, if formed with appropriate abundance, they can dominate the universe's energy density for a short duration before their evaporation. This window reveals the possibility that PBHs may have dominated the early universe before BBN and played an important role in its evolution. The consequences of an early PBH-dominated epoch have been well studied since the Hawking evaporation of PBHs involves many different and important aspects, for instance, the generation of Dark Radiation (DR)~\cite{Carr:2020xqk, Hooper:2019gtx, Lunardini:2019zob, Masina:2020xhk, Masina:2021zpu, Arbey:2021ysg}, matter-antimatter asymmetry production~\cite{Baumann:2007yr, Fujita:2014hha, Hook:2014mla, Hamada:2016jnq, Chaudhuri:2020wjo, Hooper:2020otu, Perez-Gonzalez:2020vnz, Datta:2020bht, JyotiDas:2021shi, Bhaumik:2022pil}, and the implications for the production of DM through evaporation~\cite{ Bell:1998jk, Allahverdi:2017sks, Fujita:2014hha, Lennon:2017tqq, Morrison:2018xla, Hooper:2019gtx, Masina:2020xhk, Gondolo:2020uqv, Bernal:2020kse, Bernal:2020bjf, Bernal:2020ili, Kitabayashi:2021hox, Masina:2021zpu, Cheek:2021odj,Cheek:2021cfe, Barman:2022pdo, Borah:2022iym}.

The imprints of PBHs on the induced stochastic gravitational wave background (ISGWB) can be broadly classified into a few categories according to their origin. One is from the binary mergers of the PBHs~\cite{Ali-Haimoud:2017rtz, Kohri:2018qtx, Raidal:2018bbj, Gow:2019pok, Jedamzik:2020omx, Bagui:2021dqi}, which can be produced even in the very late universe. Another class of ISGWB comes from the second-order tensor perturbations, which is related to the formation mechanism of PBHs~\cite{Saito:2008jc, Kohri:2018awv, Espinosa:2018eve, Domenech:2021ztg, Ashoorioon:2020hln, Ashoorioon:2022raz, Cai:2018dig}. 
Moreover, the gravitons produced during Hawking evaporation of PBHs can also contribute to the GWs but at a very high frequency ($10^{13}-10^{16}$ Hz), which makes this source difficult to detect~\cite{Dolgov:2011cq}. The scenario we had focused on in our earlier work \cite{Bhaumik:2022pil} is slightly different from these popular classes of PBH-GW associations. We focused on the resonant contributions~\cite{Inomata:2019ivs} of the second-order tensor perturbations from two different origins of adiabatic scalar perturbations. The first one directly comes from the inflationary adiabatic perturbations \cite{Inomata:2020lmk}, while the second is induced by the isocurvature perturbations from the PBHs population of monochromatic~\cite{ Bhaumik:2022pil,Domenech:2020ssp, Domenech:2021wkk, Papanikolaou:2020qtd} or extended mass distribution \cite{Papanikolaou:2022chm}. In this scenario, we found a distinctive doubly-peaked spectral shape of the ISGWB.

Our previous work considered only the Schwarzschild PBHs without any spin. In this work, we extend the scope of our study by including the possibility of initially spinning PBHs. We studied the case of Hawking-evaporation-driven baryogenesis in our previous work. Here we consider a broader class of possible phenomena associated with an early ultra-low mass PBH-dominated universe, connect them with the associated ISGWB signatures, and look for different effects from spinning PBHs. During their evaporation, PBHs can emit both relativistic and massive non-relativistic particles. If the emitted relativistic particles belong to the dark sector, they can contribute to the universe's total radiation energy density, usually referred to as DR. It can be characterized by an effective number of neutrinos, $\Delta N_{\rm eff}$ and can be constrained during the CMB era with current and future CMB probes or during the BBN epoch. This offers a complementary probe for our scenario besides the corresponding ISGWB. One interesting point to note here is that ISGWB also contributes to radiation energy density and, therefore, to $\Delta N_{\rm eff}$. While this ISGWB contribution to $\Delta N_{\rm eff}$ is generic to the scenario of PBH domination, the dark radiation is a special case when we consider some particular DR particles. Therefore the resulting DR can be constrained in three different channels, DR-$\Delta N_{\rm eff}$ in CMB observations, ISGWB-$\Delta N_{\rm eff}$ in CMB observations and the detection of ISGWB signals in the future GW observatories. 

Massive non-relativistic particles from Hawking evaporation can decay in a baryon-number violating process and contribute to baryogenesis. In the same context, it is possible to consider stable non-relativistic particles in the dark sector. They can constitute the observed DM energy density if they do not decay. Detection or non-detection of doubly peaked ISGWB spectrum can also very effectively constrain the parameter space for this DM production mechanism. Thus, in this work, we connect the generation of DR and DM  from PBH evaporation with corresponding ISGWB signals. We also explore whether initially spinning PBHs would leave a distinct signature, allowing us to determine the spin of these black holes. For the ISGWB sector, we find the effect of initial PBH spin to be non-zero but small. When we consider non-relativistic particles with spin $\leq 1 $ contributing to baryogenesis and DM relic density, we find a negligible effect of the initial PBH spin. The situation changes only when we consider spin-2 dark sector gravitons emission as the DR particles. In this case, we find distinguishable signatures of a non-zero initial PBH spin compared to the pure non-spinning case.

The paper is organized as follows: in section \ref{spin-pbh}, we estimate the effects of including initial PBH spins compared to the non-spinning case in the background evolution, and in section \ref{ISGWBs}, the corresponding impacts in ISGWB are discussed. We use section \ref{section4} to study the possibilities of dark radiation, dark matter, and baryogenesis for particles emitted from initially spinning PBHs. Finally, we discuss our results and implications in section \ref{discuss}. We work with $c=\hbar = k_{\rm B}=1$ and also set the reduced Planck mass $\Mpl^2=(8 \pi G)^{-1} $ to unity, unless explicitly written.
%%%%%%%%%%%%%%%%%%%%%%%%%%%%%%%%%%%%%%%%%%%%%%%%%%%%%%%%%%%%%%%%%%%%%%
%%%%%%%%%%%%%%%%%%%%%%%%%%%%%%%%%%%%%%%%%%%%%%%%%%%%%%%%
%\input{background}

\section{Spinning PBHs, their evaporation, and the background evolution}
\label{spin-pbh}

It is well known that black holes (BH) evaporate via Hawking radiation, and the emitted particles exhibit a near thermal spectrum. 
%The properties of emitted particles depend only on the specific characteristics of the BH, which, according to the no-hair conjecture, are its mass, angular momentum, and charge.
For the case of a Schwarzschild (non-rotating and uncharged) BH, the horizon temperature is given by
\begin{equation}
T^{\rm S}_{\rm PBH} = \frac{M_{\rm Pl}^2}{%8 \pi
M_{\rm PBH}} \simeq 1.053\, {\rm GeV} \left(\frac{10^{13} \, {\rm g}}{M_{\rm PBH}} \right) \, ,
\end{equation}
and the lifetime of a BH or the time scale of its complete  evaporation can be written as,
\begin{equation}
\Delta t^{\rm S}_{\rm PBH} \approx\frac{160  M_{\rm PBH}^3}{\pi~ \mathcal{G}~ \overline{g_{*,H}} ~M_{Pl}^4 } \, .
\end{equation}
Here the graybody factor $\mathcal{G} \approx 3.8$,  $g_{\star, H}$  is the number of degrees-of-freedom for particles with masses below $T_{BH}$ and $\overline{g_{*,H}}$ is average over PBH lifetime \cite{Hooper:2020otu,MacGibbon:1991tj}.
These formulas get 
modified for spinning PBHs.
For a spinning (but uncharged) PBH, the horizon temperature
is given by, 
\begin{equation}
T_{\rm PBH} = \frac{\Mpl^2}{%8 \pi
M_{\rm PBH}} \left(\frac{2\sqrt{1-a_*^2}}{1+\sqrt{1-a_*^2}}\right)\, ,
\end{equation}
where $a_*$ is the reduced spin parameter, defined by
$a_* = J{\Mpl^2}/M_{\rm PBH}^2$ and $J$ is the magnitude of the angular momentum.
The time evolution of mass and spin follow the first-order coupled differential equations,
\begin{align}
 \frac{d\Mt}{dt} &= - \varepsilon(\Mt, a(t))\frac{\MPL^4}{\Mt^2}\,, \label{BHev0}\\
 \frac{da(t)}{dt} &= - a(t)\Big[\gamma(\Mt,a(t)) - 2\varepsilon(\Mt,a(t))\Big]\frac{\MPL^4}{\Mt^3}\,.
 \label{BHev}
\end{align}
Here $\gamma$ and $\varepsilon$ are to be determined by taking into account the sum of contributions from all possible stable and unstable particles emitted due to the Hawking radiation and are functions of PBH mass $M(t)$ and spin $a(t)$ at time $t$ \cite{Page:1976df, Page:1976ki, MacGibbon:1990zk, MacGibbon:1991tj, Cheek:2022dbx}. For spinning PBHs, evaporation is more efficient, and the rate increases.
We solve equation \eqref{BHev0} and \eqref{BHev} to obtain the lifetime of a spinning PBH, $\Delta t_{\rm PBH}$, which depends both on the initial spin and mass of PBH and introduce $\mathcal{F}(a_*, M_{\rm PBH})$ as the ratio between $\Delta t_{\rm PBH}$ and the lifetime of the Schwarzschild PBH as, $\Delta t^{\rm S}_{\rm PBH}$,
\begin{align}
\label{tevp}
    \Delta t_{\rm PBH}=
      \Delta t^{\rm S}_{\rm PBH} ~\mathcal{F}(a_*, M_{\rm PBH}) \, .
\end{align}
Since the function  $\mathcal{F}(a_*, M_{\rm PBH})$ can not be calculated analytically, we estimate it  numerically using the publicly available code  \texttt{BlackHawk}
\cite{Arbey:2019mbc, Arbey:2021mbl}. We have tabulated the values of $\mathcal{F}(a_*, M_{\rm PBH})$ computed using \texttt{BlackHawk}\footnote{We find a slight mismatch between the values of $\mathcal{F}(a_*, M_{\rm PBH})$ from \texttt{BlackHawk} and \texttt{FRISBHEE} \cite{Cheek:2022dbx}. For the estimation of $\mathcal{F}(a_*, M_{\rm PBH})$ and interpolation in equation \eqref{interpol}, we only use the results from \texttt{BlackHawk}. } in Table \ref{Table:1}.   As evident from this table, for the mass range of our interest, $\mathcal{F}(a_*, M_{\rm PBH})$ bears a negligible dependence on $M_{\rm PBH}$, and we can write,
\begin{align}
    \Delta t_{\rm PBH} \approx
      \Delta t^{\rm S}_{\rm PBH} ~\mathcal{F}(a_*) \, .
      \label{intpol}
\end{align}
\begin{table}[t!]
\renewcommand{\arraystretch}{1.2}
 \begin{center}
 \begin{tabular}{|c|c|c|c|c|}
 \hline
 \text{$a_*$} & \text{$M_{\rm PBH}=10^2$ g} & \text{$M_{\rm PBH}=10^4$ g} & \text{$M_{\rm PBH}=10^6$ g} & \text{$M_{\rm PBH}=10^8$ g} \\
 \hline
 0.0 & 1.0 & 1.0 & 1.0 & 1.0 \\    
 0.1 & 0.976899 & 0.967814 & 0.972579 & 0.975514 \\     
 0.2 & 0.962462 & 0.951945 & 0.957989 & 0.961175 \\     
 0.3 & 0.938191 & 0.928076 & 0.930649 & 0.936888 \\     
 0.4 & 0.903097 & 0.894256 & 0.896733 & 0.901647 \\     
 0.5 & 0.858785 & 0.849704 & 0.852926 & 0.857398 \\     
 0.6 & 0.804028 & 0.79565 & 0.798765 & 0.802748 \\     
 0.7 & 0.736373 & 0.728771 & 0.731148 & 0.735063 \\     
 0.8 & 0.654221 & 0.64797 & 0.650297 & 0.652996 \\     
 0.9 & 0.555112 & 0.549758 & 0.55136 & 0.553933 \\     
 0.99 & 0.439075 & 0.435225 & 0.437123 & 0.438618 \\     
 0.999 & 0.421827 & 0.417918 & 0.419717 & 0.421357 \\     \hline
\end{tabular}
\vskip 10pt
\caption{We tabulate the values of the function  $\mathcal{F}(a_*, M_{\rm PBH})$ for spinning PBHs for different values of $a_*$ and for four reference values of $ M_{\rm PBH}$, computed using the code \texttt{BlackHawk}. We find that the dependence of $\mathcal{F}(a_*, M_{\rm PBH})$ on the PBH mass is  weak and thus can be neglected for the PBHs mass range of our interest.}
 \label{Table:1}
\end{center}
\end{table}
This allows us to fit the function with a polynomial of $a_*$ as
 \begin{equation}
     \mathcal{F}(a_*)=\sum_{n=0}^{4} c_n a_*^n + \mathcal{O}(a_*^5)\, ,
     \label{interpol}
 \end{equation}
where we neglect the higher-order terms as their contribution becomes insignificant. We find the values of coefficients $c_n$ as; $c_0=1.0, c_1=- 0.183014, c_2=-0.086326, c_3=-0.195741$ and $c_4=-0.110894$ to fit the numerically obtained $\mathcal{F}(a_*)$.
\begin{figure}[t]
\begin{center}
\includegraphics[scale=0.29]{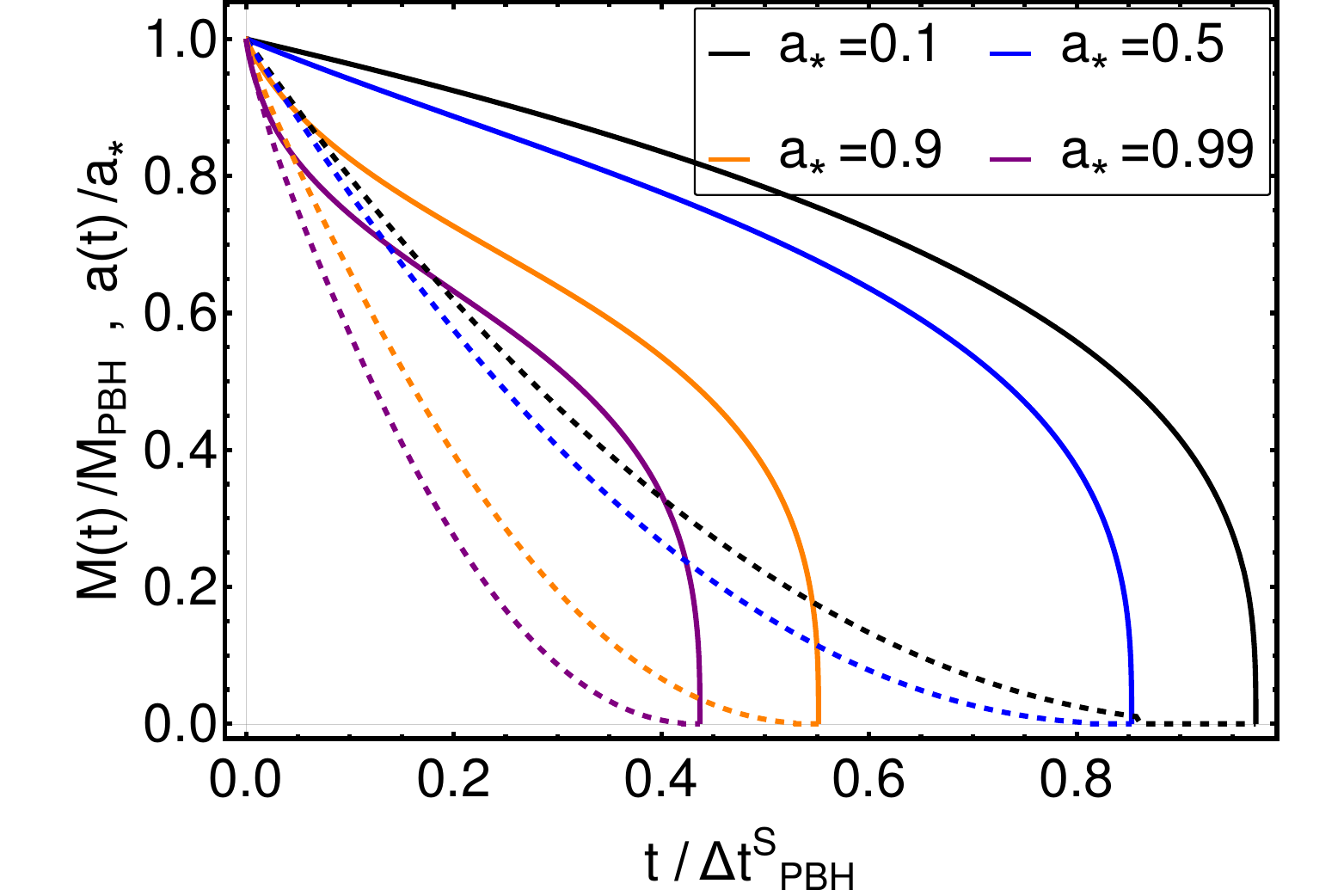}
\includegraphics[scale=0.36]{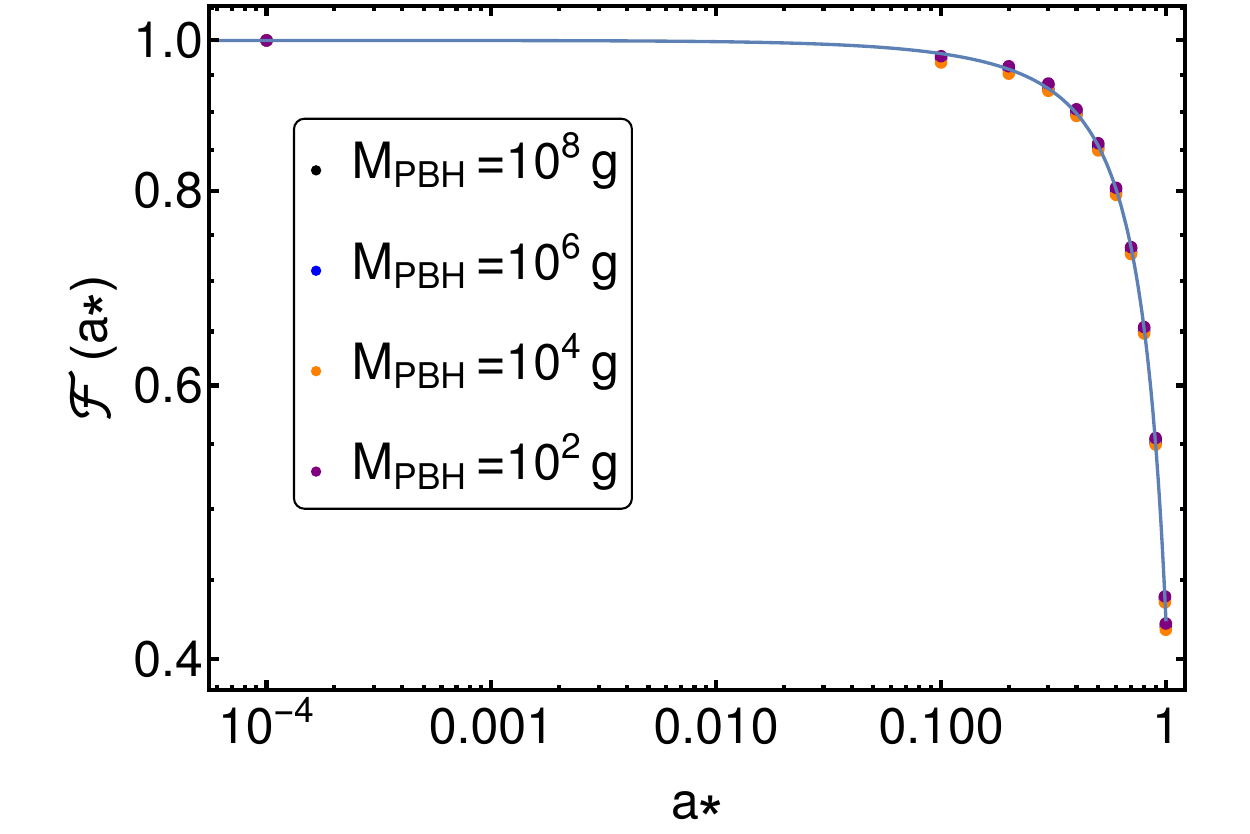}
\vskip 10pt
\caption[LoF entry] 
{ \textbf{Left panel:} The evolution of the ratio of PBHs mass $M(t)$ to their initial mass $M_{\rm PBH}$ (solid lines) and the ratio of spin parameter $a(t)$ to its initial value $a_*$ (dashed lines) as a function of time. We normalize the time axis by dividing it by the lifetime of Schwarzschild black hole of the same mass, $\Delta t_{\rm PBH}^S$. We take four different initial values of $a_{*}$ with initial PBH mass $M_{\rm PBH}=10^6$ g. 
\textbf{Right panel:} We plot $\mathcal{F}(a_*)$ as a function of initial value of PBH spin parameter $a_*$ for four 
different initial PBH masses. The data points correspond to the numerical result from \texttt{BlackHawk} while the continuous line represents a polynomial interpolation of $\mathcal{F}(a_*)$ which nicely fits the $\mathcal{F}(a_*, M_{\rm PBH})$ values obtained from the full numerical calculation. 
}
\label{sps-gw1}
\end{center}
\end{figure}
The very weak dependence of  $\mathcal{F}(a_*)$ on initial mass $M_{\rm PBH}$ is evident from the right panel of figure~\ref{sps-gw1} as the data points for different $M_{\rm PBH}$ fall on top of each other.
The left panel of Fig. \ref{sps-gw1} shows the time evolution of mass and spin of the BHs. It is clear that spin and mass evolve very differently during Hawking evaporation. While mass changes very rapidly near the end of the evaporation process, the spin depletes 
much earlier. As we go towards higher initial values of $a_*$, the lifetimes 
decrease more compared to Schwarzschild lifetime.

In our setup, we consider the ultra-low mass PBHs to form during early radiation-dominated epoch after inflation (at conformal time $\tau= \tau_f$) with initial abundance $\beta_f$. We assume that this population of PBHs dominates the universe at $\tau=\tau_m$, which lasts until their evaporation at $\tau=\tau_r$. Since we are mainly interested in calculating the resonant second-order ISGWB contribution, \cite{Inomata:2019ivs}, the PBH domination and almost instantaneous transition from matter domination to the standard radiation domination (RD) due to the Hawking evaporation play a very important role in the amplification of the ISGWB generated at the onset of RD \cite{Bhaumik:2022pil}. Using the results of our previous work \cite{Bhaumik:2022pil}, we can calculate the conformal times associated with the  PBH evaporation $\tau_{r}$ as,
 \begin{align}
\tau_{r}=\sqrt{2} \left(\frac{3\Delta t_{\rm PBH}^2\,\rho_{\rm EQ}\,\tau_{\rm EQ}^4} {\Mpl^2}\right)^{1/4}
\label{tauRD}
\end{align}
and the ratio $\tau_{\rm rat}$ between the conformal times of PBH evaporation ($\tau_{r}$) and PBH domination ($\tau_{m}$),
 \begin{align}
\tau_{\rm rat}\equiv \frac{\tau_{r}}{\tau_{m}}=2\left(\frac{3\pi^2 \Mpl^6 \,\beta_f^4 \,\tau_{r}^4 }{M_{\rm PBH}^2 \,\rho_{\rm EQ}\,\tau_{\rm EQ}^4 }    \right)^{1/6} \, .
\label{taurat}
\end{align}
Here the subscript "EQ"  refers to various quantities evaluated at the standard radiation-matter equality. As evident from eqs. (\ref{tauRD}) and (\ref{taurat}), initial non-zero values of $a_{*}$ modify the lifetime of the PBHs $\Delta t_{\rm PBH}$, and thereby affect the time duration for both PBHs domination and their evaporation. To stay within the validity of linear perturbation theory for first order scalar modes, we need to stay in the regime where $\tau_{\rm rat}  \leq 470$ ~\cite{Inomata:2019ivs, Kohri:2018awv, Assadullahi:2009nf}. We shall refer to this limit as non-linearity bound for the rest of our paper.

For each conformal time $\tau_Y$, we get a comoving wavenumber $k_Y\equiv 1/\tau_Y$, which re-enters the horizon at $\tau=\tau_Y$. It is also possible to express the values of these relevant wavenumbers in terms of  $\mathcal{F}(a_*)$. Taking $k_{\rm EQ}=1/\tau_{\rm EQ} \simeq 0.01 \text{ Mpc}^{-1}$ and $H_{\rm EQ} \simeq 20.7 \text{ Mpc}^{-1}$, we get
\begin{align}
k_{r} & = \frac{1}{\tau_r} \approx 2.1 \times 10^{11} \left[\mathcal{F}(a_*)\right]^{-1/2} \left( \frac{M_{\rm PBH}}{10^4 {\rm g} } \right)^{-3/2}  \text{Mpc}^{-1}   , \label{time1} \\
%\end{align}
%\begin{align}
 k_{m} & = \frac{\tau_{\rm rat}}{\tau_r} \approx 3.4 \times 10^{17} \left[\mathcal{F}(a_*)\right]^{-1/6} \left( \frac{M_{\rm PBH}}{10^4 {\rm g} } \right)^{-5/6} \beta_f^{2/3} \text{ Mpc}^{-1}, \label{time2} \\
%\end{align}
%\begin{align}
\hspace{2cm}  k_{f} & = \frac{k_{m}}{\beta_f}\approx3.4 \times 10^{17} \left[\mathcal{F}(a_*)\right]^{-1/6} \left( \frac{M_{\rm PBH}}{10^4 {\rm g} } \right)^{-5/6} \beta_f^{-1/3} \text{ Mpc}^{-1} .
\label{time3}
\end{align}
Note that the comoving scale $k_r$ does not depend on $\beta_f$ while both $k_m$ and $k_f$ explicitly depend on it. 
All these scales also depend upon $\mathcal{F}(a_*)$ and play a crucial role in our later analysis. 
Since all the PBHs will be evaporated by the time $\tau_r$, the comoving wavenumber $k_r$ is  associated with the transition from early matter-dominated (eMD) to the standard RD and also indicates the time of the ISGWB generation.
The scale $k_m$ points to the cutoff scale in the inflationary scalar perturbation spectrum, which leads to the first resonant peak in the ISGWB. Finally, the scale of PBH formation $k_f$ is directly related to the mass of PBHs and plays an important role in determining the cutoff scale for the PBH-induced isocurvature perturbations and, thus, the second peak of the ISGWB.

\section{Induced stochastic gravitational wave background (ISGWB) from spinning PBHs}
\label{ISGWBs}
\begin{figure}[t]
\begin{center}
\includegraphics[width=15.4cm]{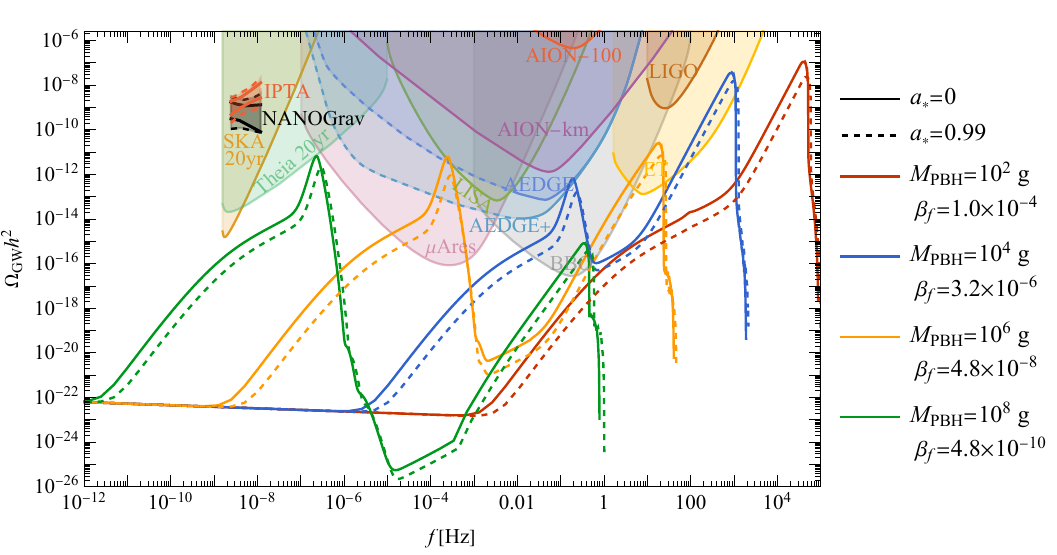}
\vskip 6pt
\caption{The spectral energy density $\Omega_{\rm GW}h^2$ of the ISGWB for selected representative populations of spinning and non-spinning PBHs with monochromatic mass range.
These populations are characterized by two parameters: $\beta_f$, the initial mass fraction of PBHs at their formation, and $M_{\rm PBH}$, the initial PBH mass.
The solid and dashed lines show our results for ISGWB for PBH populations with no spin ($a_*=0$) and with spin ($a_*=0.99$), respectively. For all these cases, the doubly peaked profiles of ISGWB  correspond to the secondary contribution sourced by the inflationary adiabatic and the isocurvature
induced adiabatic scalar perturbations, respectively.
}
\label{sps-gw2}
\end{center}
\end{figure}
\begin{figure}[t]
\begin{center}
\includegraphics[width=12.5cm, height=8cm]{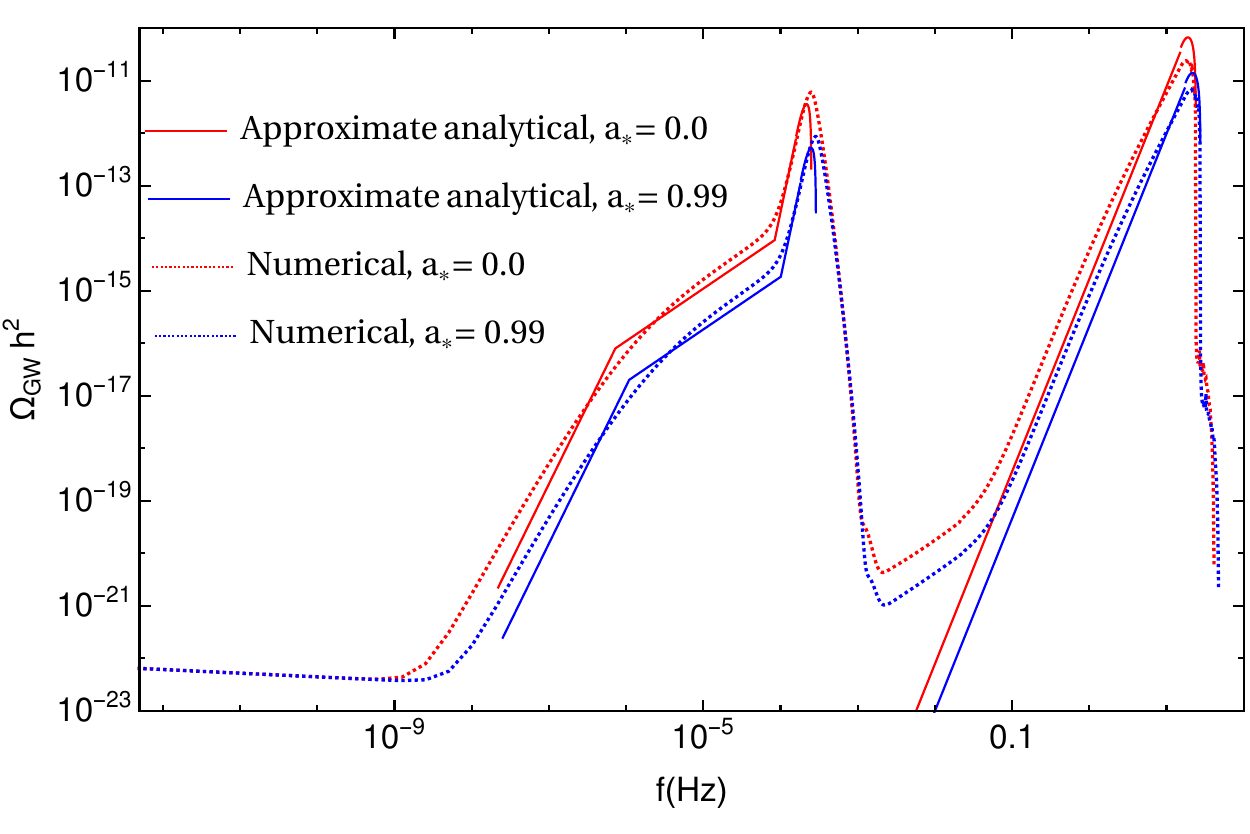}
\vskip 6pt
\caption{We compare the analytical (solid lines) and numerical (dotted lines) results for ISGWB spectral energy density $\Omega_{\rm GW}h^2$ for initial PBH mass $\MPBH =10^6 {\rm g}$ and $\beta_f=4.5\times10^{-8}$ for spinning ($a_*=0.99$) and non-spinning ($a_*=0.0$) population of PBHs. While the dotted lines are from an exact numerical calculation, the solid lines are due to the approximate analytical expressions we derived in equations \eqref{iso01} and \eqref{ad01}.}
\label{sps-gw0}
\end{center}
\end{figure}

As we discussed in the previous section, evaporating PBHs with spin have slightly different lifetimes than the non-spinning ones.
Since we focus on a scenario wherein the PBH evaporation initiates the standard RD epoch, this shift in the PBH lifetime also induces a shift in the start of the standard RD, as well as all the other relevant time scales, such as the start of PBH domination, PBH formation, etc., as shown in eqns. \eqref{time1}, \eqref{time2} and \eqref{time3}. 
%As we shall discuss later in this section, this temporal shift in the background evolution arising due to the PBHs spin also causes a similar shifting effect on the resonant peaks in the ISGWB. 

In our previous work \cite{Bhaumik:2022pil}, we estimated the resonant ISGWB from two different components of adiabatic scalar perturbations; the inflationary adiabatic and the isocurvature-induced adiabatic component. In these scenarios, the source of isocurvature perturbations is the formation of PBHs, which converts to adiabatic perturbations by the time PBHs dominate the universe.
The cutoff scale of the  initial isocurvature power spectrum, $k_{\rm UV}$, is taken at the scale of the mean distance between two black holes at their formation to avoid the granularity of the PBH fluid   \cite{Papanikolaou:2020qtd}. Taking the efficiency factor associated with PBH formation in early RD,  $\gamma \approx 0.2$ we can write,
\begin{align}
    k_{\rm UV} &=
    \left(\frac{\beta_f}{\gamma}\right)^{1/3} k_f\, .
\end{align}
The analytical expressions derived in our previous paper \cite{Bhaumik:2022pil} makes it easy to infer that the ISGWB spectrum, in this case, explicitly depends on the three wavenumbers associated with the PBH formation, domination, and evaporation; $k_f$, $k_m$, $k_r$ and the initial PBH abundance $\beta_f$. The modifications in the ISGWB due to the change in the initial PBH spin arise only through the spin dependence of these wavenumbers, as we derived in the previous section. The final form of the ISGWB spectrum today, from isocurvature contribution \cite{Bhaumik:2022pil},
\begin{align}
\Omega_{\rm GW}(\tau_0,k) & = c_g \,\Omega_{r,0} ~ \mathcal{J} \int_{-s_0}^{s_0} \frac{27 \sqrt[3]{3} \left(s^2-1\right)^2}{\left(9-3 s^2\right)^{5/3}} \dd s%\\
% & =  \frac{2}{5} c_g \,\Omega_{r,0} ~ \mathcal{J}  s_0 \left(\frac{3 \left(14-3 s_0^2\right)}{\left(1-\frac{s_0^2}{3}\right)^{2/3}}-37 \,
%   _2F_1\left(\frac{1}{2},\frac{2}{3};\frac{3}{2};\frac{s_0^2}{3}\right)\right) \, ,
   \label{iso01}
\end{align}
where $c_g \approx 0.4$ if we take the number of
relativistic degrees of freedom to be $\sim 106.7$, $\Omega_{r,0}$ is the present radiation energy density. %$_2F_1$ is the hypergeometric function
\iffalse
and,

\begin{align*}
 \hspace{1cm}    \mathcal{J}=\frac{k^3 k_{m}^8 \left(\frac{k}{k_r}\right)^{2/3}}{1327104 \sqrt[3]{2} \sqrt{3} \pi k_r^5 k_{\rm UV}^6} \, .
\end{align*}

For $k> 2/\sqrt{3}\, k_{\rm UV}$, we get a sharp cutoff in the ISGWB spectrum, but for $k< 2/\sqrt{3}\, k_{\rm UV} $, 
\fi
The limit of $s$ integral $s_0$ is defined as a function of $k$;
\begin{align}
s_0=\left\{
\begin{aligned}
&1\qquad &\tfrac{k}{k_{\rm UV}}\leq\tfrac{2}{1+\sqrt{3}}\\
&2\tfrac{k_{\rm UV}}{k}-\sqrt{3}\qquad & \tfrac{2}{1+\sqrt{3}}\leq\tfrac{k}{k_{\rm UV}}\leq\tfrac{2}{\sqrt{3}}\\
\end{aligned}
\right.\, .
\end{align}
and we can write,
\begin{align*}
 &\hspace{0cm}    \mathcal{J} \approx 2.0\times 10^{-12} \beta ^{16/3}_f \left(\frac{f}{ 1~\text{Hz}} \right)^{11/3} \left(\frac{M_{\rm PBH}}{1 g} \right)^{41/6} \mathcal{F}(a_*)^{5/2}  \, ,\\
 &\hspace{0.0cm}  \frac{k}{k_{\rm UV}} \approx 4.8\times 10^{-7}\left(\frac{M_{\rm PBH}}{1 g} \right)^{5/6} \mathcal{F}(a_*)^{1/6} \left(\frac{f}{ 1~\text{Hz}} \right) \, ,
\end{align*}
in terms of the mass of the PBHs $M_{\rm PBH}$, initial abundance $\beta_f$ and spin parameter $a_*$, after taking the shift in PBH lifetime for spinning PBHs correctly into account. This leads to a peak value for the second ISGWB peak,
\begin{align}
\Omega_{\rm GW}(\tau_0)\Big|_{k= k_{\rm UV}}\approx 1.9\times 10^{6} {\beta^{16/3}_f} \left(\frac{M_{\rm PBH}}{1 g} \right)^{34/9} \mathcal{F}(a_*)^{17/9}
\label{isopeak}
\end{align}
The first or the low-frequency peak of the ISGWB spectrum corresponds to the resonant contribution from inflationary adiabatic perturbations. We consider the standard power-law power spectra for inflationary scalar curvature perturbations, 
\begin{align}
\label{power-law}
\mathcal{P}_{\mathcal{R}}=A_s \left( \frac{k}{k_{p}} \right)^{n_s -1} \, ,
\end{align}
with pivot scale $k_p=0.05 \, {\rm Mpc}^{-1}$, scalar amplitude $A_s= 2.09 \times 10^{-9} $ and scalar index $n_s=0.965$ \cite{Planck:2018jri}, which leads to the first peak of ISGWB \cite{Bhaumik:2022pil, Inomata:2019ivs},
\begin{align}
\frac{\Omega_{\rm GW}(\tau_0,k) }{A_{\text{s}}^2 c_g \,\Omega_{r,0}} \simeq & 
\begin{cases}
3 \times 10^{-7} x_r^3 x_{\text{max}}^5 \qquad &   150 x_{\text{max}}^{-5/3} \lesssim x_r \ll 1 \\
6.6 \times 10^{-7} x_r x_{\text{max}}^5 & 1 \ll x_r \lesssim x_{\text{max}}^{5/6} \\
3 \times 10^{-7}   x_r^7 &     x_{\text{max}}^{5/6} \lesssim x_r   \lesssim \tfrac{2}{1+\sqrt{3}}x_{\text{max}}   \\
\mathcal{C}(k)  &  \tfrac{2}{1+\sqrt{3}}\leq\tfrac{x_r}{x_{\text{max}}}\leq\tfrac{2}{\sqrt{3}}\\
 \end{cases} \, ,
 \label{ad01}
\end{align}
where 
\begin{eqnarray}
 \mathcal{C}(k) =0.00638 \times 2^{-2 n_s-13} ~3^{n_s} ~x_r^7 ~s_0
\left(\frac{x_r}{x_{\rm max}}\right)^{2 n_s-2} \times \hspace*{5cm}\nonumber \\
 \left[-s_0^2 \, _2F_1\left(\frac{3}{2},-n_s;\frac{5}{2};\frac{s_0^2}{3}\right)   +4 \, _2F_1\left(\frac{1}{2},1-n_s;\frac{3}{2};\frac{s_0^2}{3}\right)-3 \,  _2F_1\left(\frac{1}{2},-n_s;\frac{3}{2};\frac{s_0^2}{3}\right) \right] \, .
\end{eqnarray}
%Here $n_s \approx  0.964$ refers to the scalar spectral index for the nearly scale-invariant power spectrum of the inflationary adiabatic part. 
and,
\begin{align*}
 &  x_r =k/k_r \approx 0.001\left(\frac{f}{ 1~\text{Hz}} \right) \left(\frac{M_{\rm PBH}}{1 g} \right)^{3/2} \sqrt{\mathcal{F}(a_*)} \, , \\
& x_{max} =k_{m}/k_r \approx 2.36*10^3 \beta_f^{2/3} \left(\frac{M_{\rm PBH}}{1 g} \right)^{2/3} {\mathcal{F}(a_*)}^{1/3} \, , \\
& s_0 = 2\frac{k_{m}}{k}-\sqrt{3} \approx 1.16*10^{-6}\left(\frac{f}{ 1~\text{Hz}} \right)\beta_f^{-2/3} \left(\frac{M_{\rm PBH}}{1 g} \right)^{5/6} {\mathcal{F}(a_*)}^{1/3} -\sqrt{3} \, .
\end{align*}
The peak of the inflationary adiabatic contribution for ISGWB is expected to appear at $k=k_m$, and turns out to be,
\begin{align}
\Omega_{\rm GW}(\tau_0)\Big|_{k= k_{ m}}\approx 6.9\times 10^{-6} {\beta_f}^{14/3} \left(\frac{M_{\rm PBH}}{1 g} \right)^{14/3} \mathcal{F}(a_*)^{7/3} \\
\approx 1.7 \times 10^{-29} \left( \frac{k_m}{k_r} \right)^7 \equiv 1.7 \times 10^{-29} \left( \frac{\tau_r}{\tau_m} \right)^7 \, .
\label{adpeak}
\end{align}
To stay within the validity of linear perturbation theory for scalar modes, we restrict our study only in the regime where $\left( \frac{\tau_r}{\tau_m} \right) \leq 470$ ~\cite{Inomata:2019ivs, Kohri:2018awv, Assadullahi:2009nf}, which puts a strict bound on the inflationary adiabatic peak, 
\begin{equation}
    \Omega_{\rm GW}(\tau_0)\Big|_{k= k_{ m}} \leq 8.4 \times 10^{-11} \, .
    \label{adlimit}
\end{equation}
To check the viability of analytical results of equation \eqref{iso01} and \eqref{ad01}, we plot it along with numerically obtained ISGWB spectral energy density $\Omega_{\rm GW}h^2$ in Fig. \ref{sps-gw0} both for spinning and non-spinning PBHs. As we discussed in section \ref{spin-pbh}, we use $\mathcal{F}(a_*, \MPBH)$, obtained from \texttt{BlackHawk} to numerically compute the ISGWB spectrum for spinning PBHs. In Fig. \ref{sps-gw2}, we plot the ISGWB spectra for PBH mass range $10^2-10^8$ g with appropriate mass fractions, both for initially spinning and non-spinning PBHs,
along with the projected  sensitivities~\cite{Thrane:2013oya} of LIGO~\cite{LIGOScientific:2014pky,LIGOScientific:2016fpe},
 SKA~\cite{Janssen:2014dka}, LISA~\cite{Bartolo:2016ami,Auclair:2022lcg},  AEDGE~\cite{Badurina:2021rgt,AEDGE:2019nxb}, AION/MAGIS~\cite{Badurina:2021rgt,Badurina:2019hst,Graham:2016plp,Graham:2017pmn}, ET~\cite{Punturo:2010zz,Hild:2010id}, BBO~\cite{Yagi:2011wg,Crowder:2005nr},
 $\mu$ARES\cite{Sesana:2019vho},
 and Theia\cite{Garcia-Bellido:2021zgu}. 
 We also show fits to the possible signal reported by PTA collaborations~\cite{NANOGrav:2020bcs,Goncharov:2021oub,Chen:2021rqp,Antoniadis:2022pcn}. Both from Figs. \ref{sps-gw2} and \ref{sps-gw0}, weak dependence of ISGWB on the initial PBH spin is quite evident. 
 
In this work, we are primarily interested in the domination and evaporation signatures of ultra-light PBHs and we remain agnostic about their generation mechanism. As the simplest possible case, we take both the initial mass and spin distributions to be monochromatic. However, a broader distribution for PBH spin and mass can be more attractive in the context of a specific PBH formation mechanism. For broader distribution of either PBH mass or spin would lead to a longer duration of the transition from PBH domination to RD phase and, therefore, will suppress the resonant ISGWB peaks for both isocurvature-induced and inflationary adiabatic peaks. As the suppression effects are more dominant for higher wavenumber modes \cite{Inomata:2020lmk}, we can expect that the isocurvature peak will be more strongly affected than the inflationary one. Recently, effects of broader PBH mass distributions have been studied in the context of isocurvature-induced ISGWB peak \cite{Papanikolaou:2022chm}. 
Since an accurate estimation of the ISGWB for a broader spin or mass distribution requires an integrated setup to carefully keep track of the adiabatic and isocurvature scalar modes and second-order tensor modes during the transition, we leave this analysis for future work.

%As the simplest possible case,  here we confine ourselves to the scenario where both initial mass and spin distributions are taken to be monochromatic. A broader distribution for PBH spin and mass can be more attractive in the context PBH formation mechanism. Broader distribution of either PBH mass or spin would lead to a longer transition phase from PBH domination to radiation-dominated phase and therefore suppress the resonant ISGWB peaks for both isocurvature-induced and inflationary adiabatic peaks. As the suppression effects are more dominant for higher wavenumber modes \cite{Inomata:2020lmk}, we can predict the isocurvature peak to get more strongly affected than the inflationary one. Recently effects of the broader mass distribution have been studied in the context of isocurvature-induced ISGWB peak \cite{Papanikolaou:2022chm}. But as an accurate estimation of induced gravitational wave background in such a scenario requires an integrated setup to carefully keep track of scalar and second-order tensor modes during the transition, I leave this study for broader spin distribution for future work. 

\section{Signatures of spinning PBHs in different scenarios}
\label{section4}

The Hawking evaporation of PBHs takes place via the emission of all kinds of particles depending on the properties of the emitted particles and the mass and angular momentum of the evaporating PBH. In this process, the mass and angular momentum of a PBH is dissipated, which depends on its initial properties.
The rate of emission for a particular particle species  $i$, with spin $s_i$ in the energy interval $(E, E+dE)$ can be expressed as \cite{Masina:2021zpu,Arbey:2021mbl,Arbey:2019mbc}, 
\be
 \frac{d^2N_i}{dt \,dE} 
=  \frac{g_i}{2 \pi } \sum_{\ell,m} \Gamma_{s_i \ell m}\big[E,M_{\rm BH}(t),a_*(t)\big]\frac{1}{e^{(E_i-m \Omega)/{T_{\rm BH}(t)}} -(- 1)^{2 s_i}} ,
 \label{blackhawk-primary}
 \ee
  with the total energy $E^2= p^2 c^2+ m_i^2 c^4$, particle mass $m_i$, and projection to the angular momentum $m \in [-\ell, +\ell]$,
$\Omega= (4 \pi/M_{\rm BH})(a_*/(1+\sqrt{1-a^2_*}))$  is the angular velocity of the horizon and internal  degrees of freedom (DOF) $g_i$ accounts for the polarization and color DOF.   
The graybody factors $\Gamma_{s_i \ell m}$ characterize the probabilities for the emitted species not to be re-absorbed, escaping the gravitational well of the BH, which are a function of  $E$, $M_{BH}(t)$, $a_*(t)$, $g_i$, $s_i$ and $m_i$.

Different particles emitted from the PBHs can contribute to very different physical phenomena. Massive non-relativistic particles emitted from PBHs can contribute to the observed matter anti-matter asymmetry in our universe through their baryon number violating decay, which we considered in our previous work \cite{Bhaumik:2022pil}. On the other hand, Hawking-production of massive non-relativistic stable dark-sector particles can account for our universe's total dark matter budget while the production of dark sector light relativistic particles can contribute to the dark radiation \cite{Hooper:2019gtx}.

\subsection{Dark radiation}
\label{darkradiation}

The relativistic dark sector particles emitted from Hawking radiation of PBHs would add to the total radiation energy density. Since the background dynamics are  probed very precisely both during CMB and BBN eras, this extra dark radiation (DR) component can be constrained. Although the effects of a broader spin distribution in the context of DR have been studied earlier \cite{Arbey:2021ysg}, similar to the previous section, we restrict our analysis to the case of a monochromatic mass and spin distribution of PBHs, while estimating the DR contribution in this section.

We usually parameterize the presence of this extra radiation in terms of the effective number of relativistic degrees of freedom, $\Delta N_{\rm eff}$. The total radiation energy density can be written as a sum of the bosonic and fermionic components,
\be
\rho_{\rm rad} =
\frac{\pi^2}{30}
\left[
{\sum_b}~g_{*b}
\left(\frac{T_b}{T}\right)^4 
+{\frac{7}{8}}\,{\sum_f}~g_{*f}\left(\frac{T_f}{T}\right)^4
\right]\,T^4 \, .
\label{eqneff}	
\ee
After neutrino decoupling, when the temperature drops below the electron mass, electron-positron annihilates together to produce photons. Thus the entropy of the electrons and positrons is transferred to the photons but not to the decoupled neutrinos. This leads to a relative difference between photon and neutrino temperatures,
\be
T_\nu = (4/11)^{1/3}T_\gamma\, ,
\ee
and using \eqref{eqneff} we have,
\be
\rho_\mathrm{R}=\rho_\gamma\left[1+\frac{7}{8}\left(\frac{4}{11}\right)^{4/3}N_\mathrm{eff}\right]\,.
\ee
In the presence of an additional dark radiation component from PBHs evaporation, we can express it with,
\be
\rho_\mathrm{R} + \rho_\mathrm{DR}=\rho_\gamma\left[1+\frac{7}{8}\left(\frac{4}{11}\right)^{4/3} \left(N_\mathrm{eff}^{\rm SM} + \Delta N_\mathrm{eff}   \right)\right] \, .
\label{eqNeff1}
\ee
At matter-radiation equality ($\tau=\tau_{EQ}$), we can therefore, write  
 $\Delta N_\mathrm{eff}$  as
 \be
 \Delta N_{\rm eff}\Big|_{\rm DR} = \left\{\frac{8}{7}\left(\frac{4}{11}\right)^{-\frac{4}{3}}+N_{\rm eff}^{\rm SM}\right\} 
 \frac{\rho_{\rm DR }(\tau_{\rm EQ})}{\rho_{\rm R}(\tau_{\rm EQ})}\, ,
\label{eqNeff2}
\ee
and it is straightforward to connect it with the quantities at the time of PBH evaporation ($\tau=\tau_{r}$),
 \be
 \frac{\rho_{\rm DR }(\tau_{\rm EQ})}{\rho_{\rm R}(\tau_{\rm EQ})}=
 \frac{\rho_{\rm DR}(\tau_{\rm r})}{\rho_{\rm R}^{\rm SM}(\tau_{\rm r})}
 \left(\frac{g_*(T_{\rm r})}{g_*(T_{\rm EQ})}\right)
 \left(\frac{g_{*S}(T_{\rm EQ})}{g_{*S}(T_{\rm r})}\right)^{\frac{4}{3}}\, .
\ee
\begin{figure}[t]
\begin{center}
\includegraphics[width=15cm]{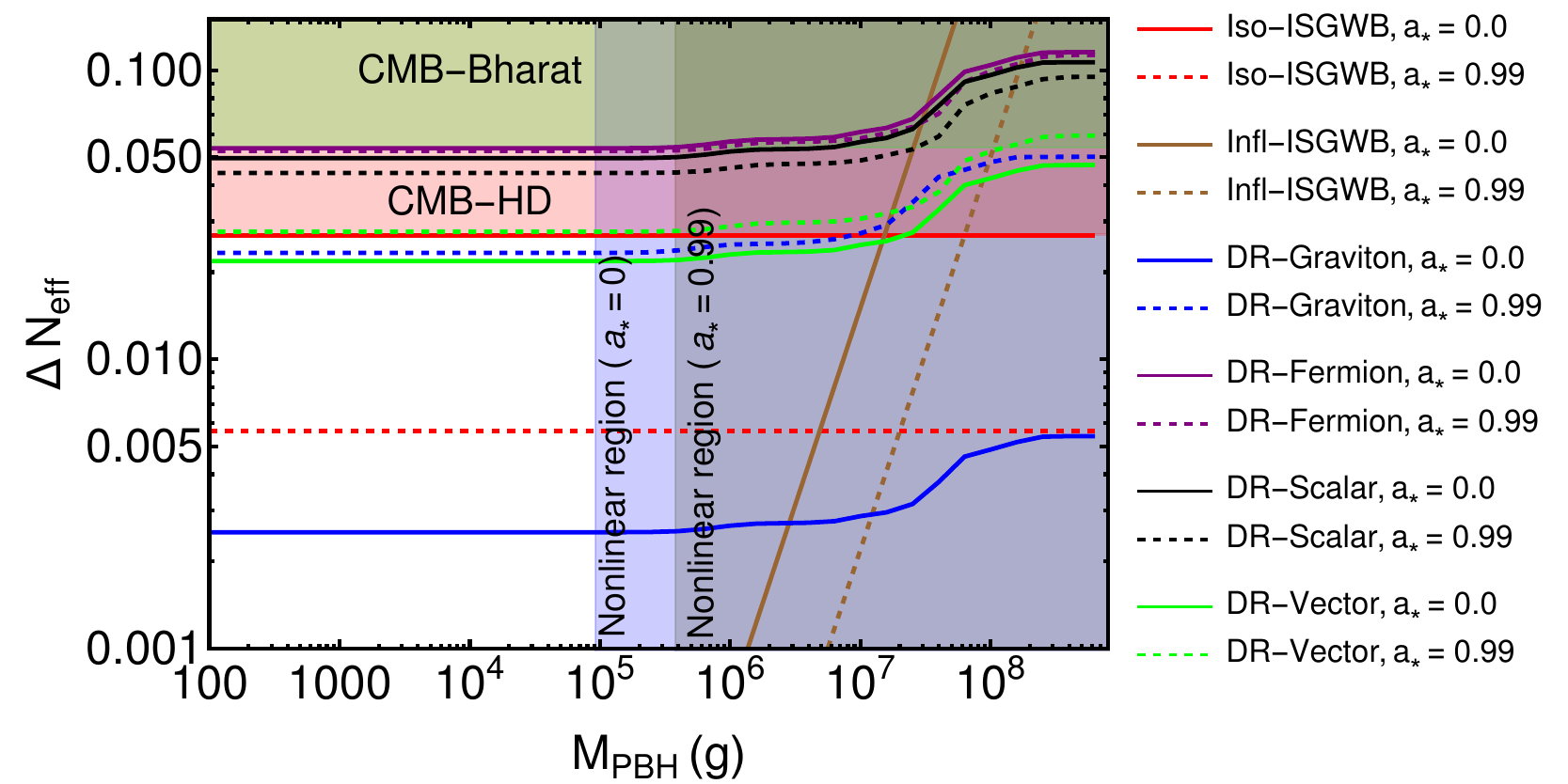}
\vskip 10pt
\caption{ Comparison of different contributions to $\Delta N _{\rm eff}$, while we change the initial abundance of PBHs with changing PBH mass: $\beta_f \propto 1/M_{\rm PBH}^{17/24}$ to keep $\Delta N _{\rm eff}$ contribution from isocurvature induced adiabatic peak of ISGWB (red line) to a constant value and plot corresponding $\Delta N _{\rm eff}$ contribution considering different massless dark radiation particles. In these cases, solid lines correspond to $a_*=0.0$, and dashed lines refer to  $a_*=0.99$. Also, for all these cases, we maintain $\beta_f \left(\frac{ M_{\rm PBH}}{1.0 \texttt{ g} }\right)^{17/24}={\rm constant}$ and choose this constant appropriately such that the isocurvature induced ISGWB contribution to $\Delta N _{\rm eff}$ is comparable. The non-linear regime ($\tau_{\rm rat} \ge 470$) is shaded in grey, and the CMB-HD and CMB-Bharat sensitivities are displayed with the  pink and green shaded regions. We also get detectable $\Delta N _{\rm eff}$ from ISGWB peak of inflationary adiabatic perturbation (brown line) only in the non-linear region, as expected from equation \eqref{adpeak} and \eqref{adlimit}.}
\label{DRR1}
\end{center}
\end{figure}

\begin{figure}[t]
\begin{center}
\includegraphics[width=15cm]{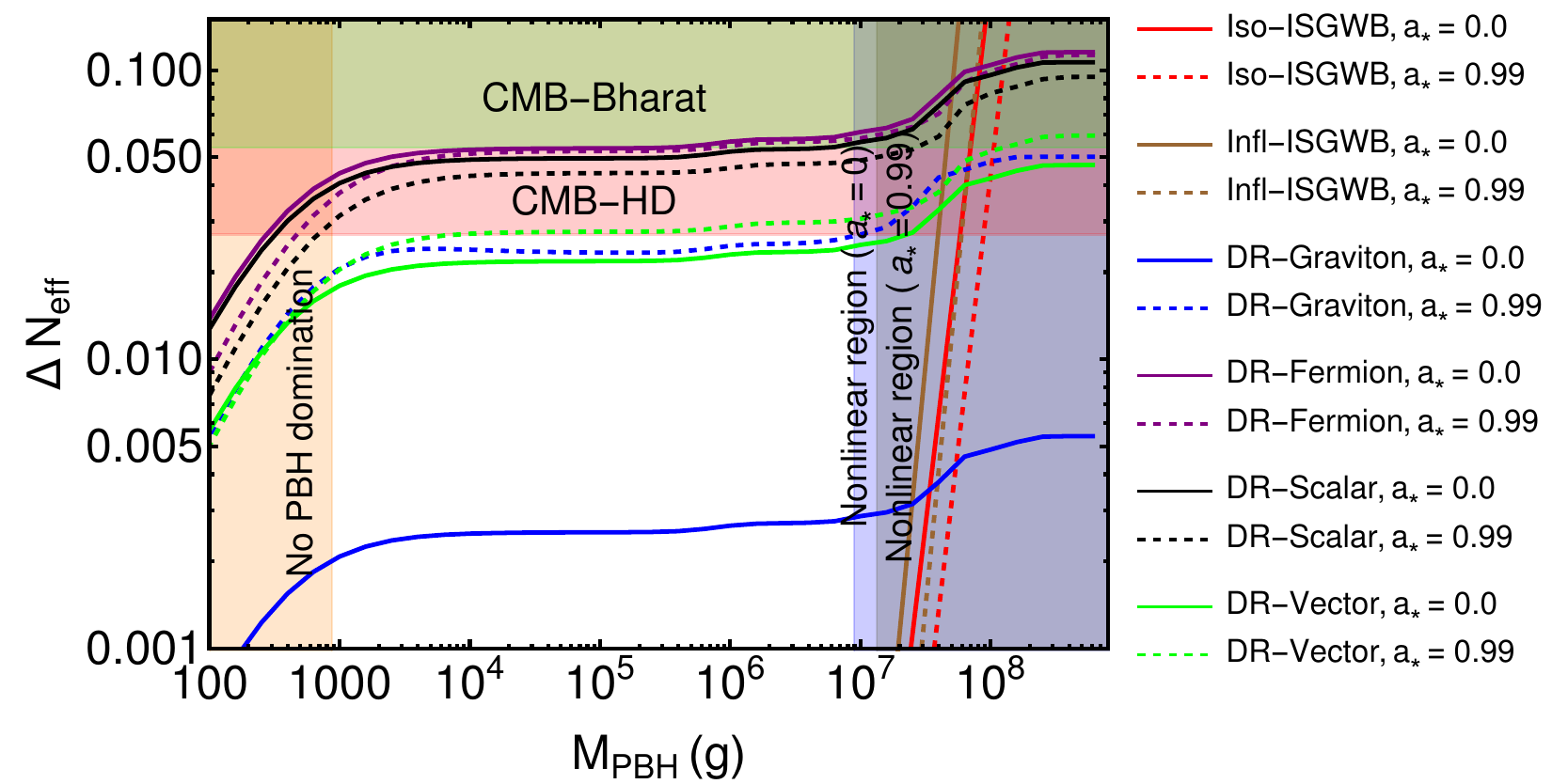}
\vskip 10pt\caption{ Comparison of different contributions to $\Delta N _{\rm eff}$, while we keep the initial abundance of PBHs fixed $\beta_f =10^{-8}$ and plot the corresponding $\Delta N _{\rm eff}$ contribution from the isocurvature induced adiabatic peak of ISGWB (red line) and different massless DR particles. For each of these cases, solid lines correspond to $a_*=0.0$, and dashed lines refer to $a_*=0.99$. The non-linear regime is shaded in grey, the CMB-HD and CMB-Bharat sensitivities with the pink and green shaded regions, and the parameter space of no PBH domination in which they evaporate before they can dominate, in orange.  We also plot $\Delta N _{\rm eff}$ from an inflationary adiabatic peak of ISGWB (brown line) and find it detectable only in the non-linear region.}
\label{DRR2}
\end{center}
\end{figure}

\begin{figure}[t]
\begin{center}
\includegraphics[width=13cm,height=7.6cm]{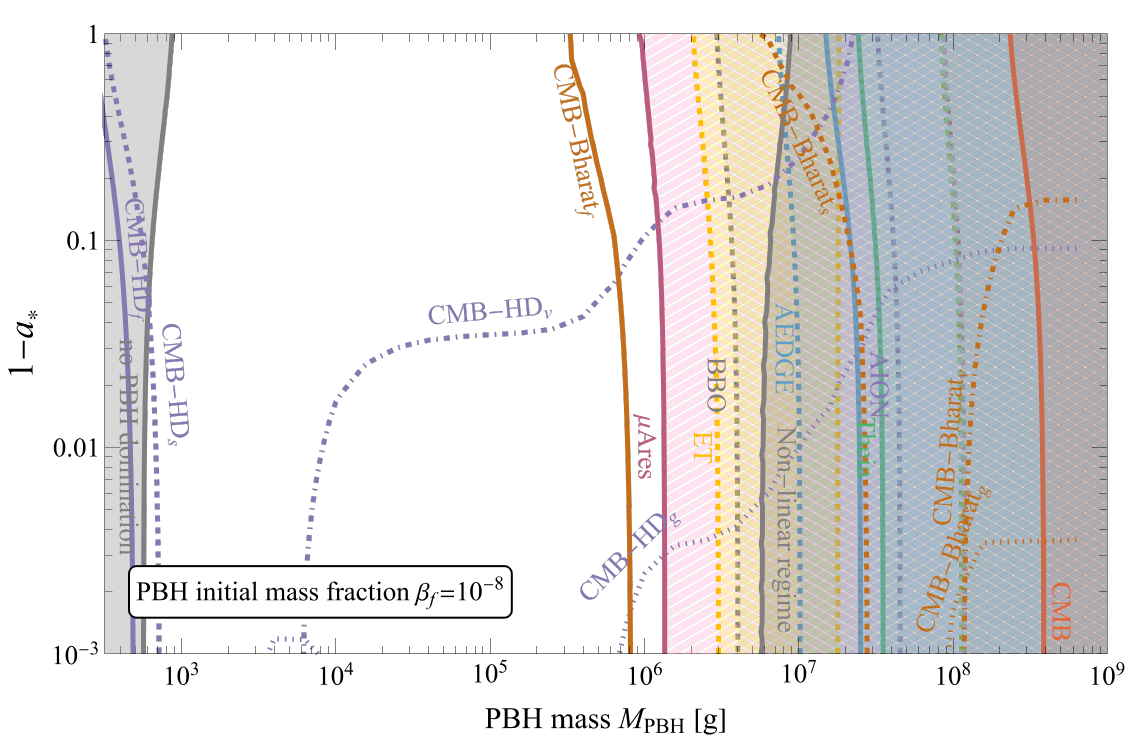}
\vskip 4pt
\includegraphics[width=13cm,height=7.6cm]{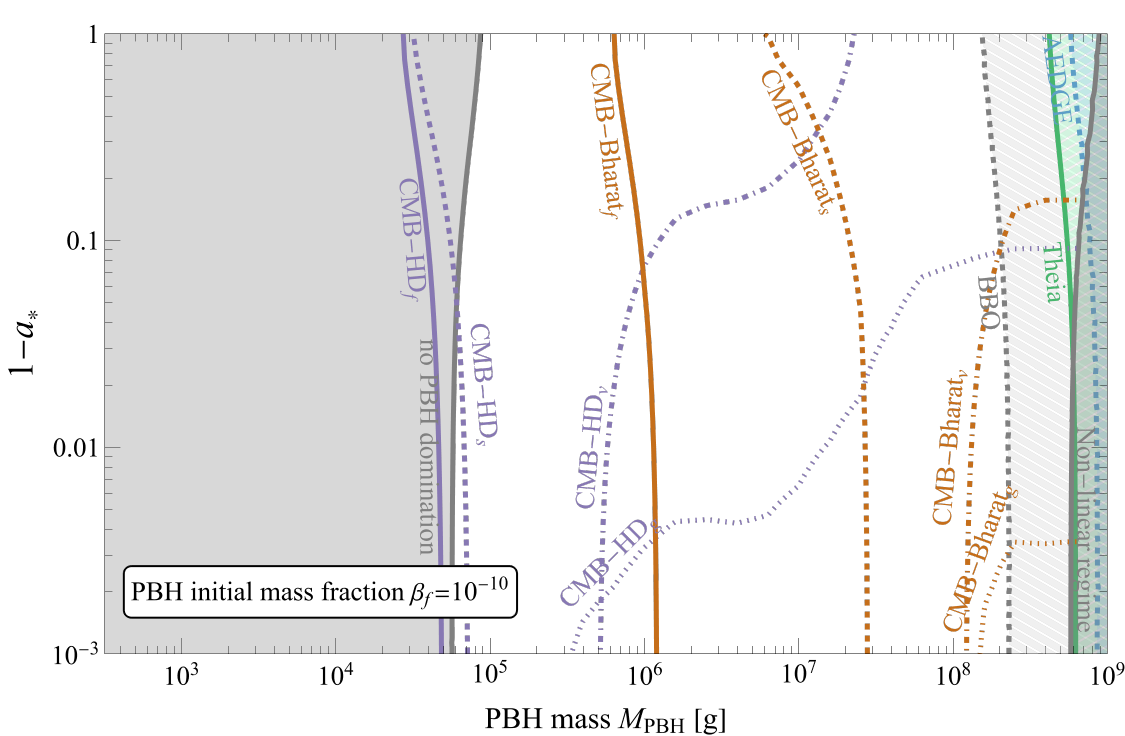}
\vskip 10pt
\caption{The various contours display the reach of the indicated GW observatories and future CMB experiments due to the produced DR for two different choices of the PBH initial mass fraction $\beta_f$. The detection of each of our two peaks is indicated in a separate way for each experiment: low-frequency inflationary adiabatic contribution detection is shown by '$/$' dashed filling and solid contours, while the high-frequency isocurvature-induced adiabatic peak is indicated by '$\backslash$' dashed filling and dashed contours.  We assume one extra scalar, fermionic, vector, or graviton particle indicated by the subscripts $s$,$f$,$v$, or $g$ to be responsible for DR production. For the lines denoted by CMB-HD or CMB-Bharat, the higher $\MPBH$ and smaller $1 - a_*$ region denotes the parameter-space where $\Delta N_{\rm eff}$ from DR is detectable in these experiments.
%The subscripts $s$,$f$,$v$, and $g$ denote scalar, fermionic, vector, and graviton particles, respectively. %The horizontal axis indicates the PBH mass, while the vertical axis shows the initial spin of the PBHs. 
The grey region on the left corresponds to the `No PBH domination' while the grey region on the right indicates the `Non-linear regime'. %described by $\tau_{\rm rat} \leq 470$. %Both these regions depend crucially on the value of $\beta_f$. 
}
\label{spincontours}
\end{center}
\end{figure}

While calculating the energy density of dark radiation particles at the time of PBH evaporation $\rho_{\rm DR}(\tau_{\rm r})$, the approximation of instantaneous production of these particles at the very end of PBH evaporation process is quite justified for non-spinning PBHs and for the production of particles with spin $\le 1$ even for spinning PBHs \cite{Hooper:2019gtx, Hooper:2020evu,Masina:2020xhk, Masina:2021zpu}. However, for the spin $2$ particles produced from initially spinning PBHs to act as dark radiation particles, this approximation breaks down \cite{Cheek:2022dbx}. The light relativistic particles are produced in the earlier phase of PBH evaporation compared to the massive particles.
%In this particular scenario, the higher spin particles are produced abundantly at the initial phase of PBH evaporation, while the other spin $\le$ 1 particles are produced later. 
Thus, by the time the entire PBH evaporates, the initial energy density of these particles dilutes significantly. This effect is more important for PBH domination scenarios because in such a scenario, while the total energy density will dilute as $a^{-3}$, the energy density of light relativistic particles would dilute as $a^{-4}$. This effect was pointed out and taken into account by solving the Boltzmann equations for combined fluid composed of SM radiation, dark radiation, and PBH in a publicly available code \texttt{FRISBHEE} ~\cite{Cheek:2022dbx}. We also use this code for our set-up to calculate $\rho_{\rm DR}(\tau_{\rm r})$ taking into account the redshift effects and to estimate $\Delta N_{\rm eff}$ accurately.

 As we mentioned, ISGWB also contributes to the radiation energy density and, therefore, to $\Delta N_{\rm eff}$. It is straightforward to generalise equation \eqref{eqNeff2} for ISGWB contribution,
  \be
 \Delta N_{\rm eff}\Big|_{\rm ISGWB}= \left\{\frac{8}{7}\left(\frac{4}{11}\right)^{-\frac{4}{3}}+N_{\rm eff}^{\rm SM}\right\} 
 \frac{\rho_{\rm GW }(\tau_{\rm EQ})}{\rho_{\rm R}(\tau_{\rm EQ})}= \left\{\frac{8}{7}\left(\frac{4}{11}\right)^{-\frac{4}{3}}+N_{\rm eff}^{\rm SM}\right\} 
 \frac{\Omega_{\rm GW }(\tau_{0})}{\Omega_{\rm R}(\tau_{0})}\, .
\ee
The non-linearity bound $\tau_{\rm rat} \leq 470 $ puts a strong constraint on the inflationary adiabatic peak of 
 the ISGWB and this limit in equation \eqref{adlimit} translates to,
\be 
\Delta N_{\rm eff}\Big|_{\rm infl-ISGWB} \leq 2.5 \times 10^{-5} \,
\ee
which is too small to be probed by BBN or any recent or upcoming CMB observatories. It can also be seen in Fig.~\ref{DRR2} and Fig.~\ref{DRR1} that $\Delta N_{\rm eff}$ from the first peak of ISGWB, is only detectable in the non-linear region and, therefore, is not of our interest.

That only leaves the isocurvature-induced adiabatic contribution of ISGWB to be considered as a candidate  for significant contribution to $\Delta N_{\rm eff}$.
From \eqref{isopeak}, we can write,
\be
\Delta N_{\rm eff}\Big|_{\rm Iso-ISGWB} \simeq 5.71\times10^{11} \beta_f^{16/3} \left(\frac{M_{\rm PBH}}{1 g} \right)^{34/9} {\mathcal{F}(a_*)}^{17/9} \,
\label{DNeffGW}
\ee
where we take $N_{\rm eff}^{\rm SM} \simeq 3.048$ and $\Omega_{\rm R}(\tau_{0}) h^2 \simeq 2.5 \times 10^{-5}$ \cite{Planck:2018vyg}.
From equation \eqref{DNeffGW}, it is evident that the dependence of PBH mass and abundance to $\Delta N_{\rm eff}$ comes from $ \beta_f^{16/3} M_{\rm PBH}^{34/9}$ factor for ISGWB contribution. We plot the corresponding DR components keeping this factor to a constant value in Fig.~\ref{DRR1}. As a result, we obtain the $\Delta N_{\rm eff}$ from the isocurvature peak for spinning and non-spinning PBHs, as different constant values. ${\mathcal{F}(a_*)}$ terms contribute to the difference between spinning and non-spinning PBH cases.

In Fig.~\ref{DRR2}, we keep the initial PBH abundance $\beta_f$ fixed and show corresponding $\Delta N_{\rm eff}$ contributions for ISGWB and DR. As ISGWB peaks are strongly dependent on $\MPBH$, we get highly tilted lines for ISGWB contributions. The DR contribution reflects a somewhat different nature: for higher mass PBHs, DR contribution is the same as in Fig. \ref{DRR1} and Fig. \ref{DRR2}, but for lower mass PBHs, constant $\beta_f$ results in a suppressed contribution to $\Delta N_{\rm eff}$ in Fig. \ref{DRR2}, for the region where PBHs evaporate before they can dominate. As we go to even lower mass PBHs, they evaporate earlier, and the fraction of PBHs and the fraction of DR particles emitted from PBHs  decreases. This effect leads to suppression in $\Delta N_{\rm eff}$ from DR, in the left part of Fig. \ref{DRR2}. We can also see from Fig. \ref{DRR1} and Fig. \ref{DRR2} that as long as $\beta_f$ is large enough to ensure that PBHs dominate the universe before they evaporate, increasing the value of $\beta_f$ does not lead to a significant difference in $\Delta N_{\rm eff}$.

This difference in nature of $\Delta N_{\rm eff}$ from ISGWB and DR suggests a novel complementarity. While $\Delta N_{\rm eff}$ from DR is not very sensitive to either the initial abundance of PBHs $\beta_f$ or PBH mass $M_{\rm PBH}$ and can only differentiate whether there was a PBH domination epoch or not, $\Delta N_{\rm eff}$ from ISGWB has a razor-sharp sensitivity with changing values of $\beta_f$ and $M_{\rm PBH}$. For the DR contributions, we consider massless particles with spin $s=0,1/2,1$ and 2. As the PBH spin increases, the distribution of emitted particles favors higher spin particles. We can see this trend from Fig. \ref{DRR1} and \ref{DRR2}. For the scalar or $s=0$ particles, spinning PBHs contribute less to $\Delta N_{\rm eff}$ than the  Schwarzschild case, but as we go towards higher spin particles, the contribution from spinning PBH increases. For fermions with $s=1/2$, spinning and non-spinning PBHs contribute almost equally; for $s=1$ vector particles, spinning PBHs contribute slightly more than non-spinning ones, but this difference becomes visibly large for $s=2$ graviton particles. For gravitons, the spinning PBHs produce an order of magnitude higher value of $\Delta N_{\rm eff}$ compared to the Schwarzschild case. 
%Recently the calculation of DR has been improved, taking into account the dilution effects for graviton energy density as they emit predominantly during the early phase of PBH evaporation \cite{Cheek:2022dbx}. We also use \texttt{FRISBHEE} to include the resulting suppression from the dilution effect. 

It is evident from Figs. \ref{DRR1} and \ref{DRR2}, that while the scalar and fermion contributions to  $\Delta N_{\rm eff}$ are relevant for the projected sensitivities of both CMB-HD \cite{CMB-HD:2022bsz} and CMB-Bharat \cite{CMBbharat:01} experiments, the graviton, and vector contributions can only be probed by CMB-HD experiment. We do not show the sensitivity of any other future CMB experiments like CMB-S4 \cite{CMB-S4:2022ght} as they are not sensitive to our expected $\Delta N_{\rm eff}$ contribution from DR. As the graviton contribution shows a strong dependence on the initial PBH spin, we can obtain a PBH mass dependent critical value for $a_*$, only above which we can expect associated $\Delta N_{\rm eff}$ to be detectable in CMB-HD experiment. 

In Fig. \ref{spincontours}, we show the effects of changing PBH spin in a more detailed manner and connect detection sensitivities for different GW detectors with existing and projected CMB observational bounds on $\Delta N_{\rm eff}$ both for DR and ISGWB contributions.
Similar to our earlier work \cite{Bhaumik:2022pil}, we set the detection criteria in terms of signal-to-noise ratio (SNR),
\be
{\rm SNR} \equiv \sqrt{\mathcal{T}\int {\rm d}f\, \left[\frac{\Omega_{\rm GW}(f)}{\Omega_{\rm noise}(f)}\right]^2} \ge 10 \, .
\ee
Assuming the operation time $\mathcal{T}=4$~yr for each experiment, we calculate the SNR by using the noise curve of a given experiment and display the results for the entire parameter space of interest in figure \ref{spincontours}. We also show  (the solid red line in Fig.  \ref{spincontours}) the strong upper bound on the parameter space coming from the overproduction of GWs spoiling the CMB~\cite{Henrot-Versille:2014jua, Smith:2006nka}.

\iffalse
\ag{In addition to the measurements of dark radiation, there is a long-standing problem with the Hubble value measurements from CMB data and late-time supernova data, and having the presence of dark radiation provides a solution to such an anomaly, as described in detail in Refs. \cite{Anchordoqui:2020djl,Vagnozzi:2019ezj}. Using the relation 
\be
H_0 = H_{\rm CMB} + 6.2 \Delta N_{\rm eff}
\ee
with $H_{\rm CMB} = 67.9$ km/s/Mpc.
For $\Delta N_{\rm eff} \sim 0.7 $, one can see that CMB and late time measurements of the Hubble constant would reconcile at 1 $\sigma$ level.
}
\ag{We could show a plot between $H_{0}$ versus PBH mass, maybe since PBH mass contributes to Neff ?}
\fi

\subsection{Dark matter relic density}

PBH evaporation can lead to the non-thermal production of stable heavy DM particles. We can also consider light mass DM particles from Hawking evaporation, but as they are highly constrained from particle physics experiments, instead, we focus on heavier mass DM particles in the form of Dirac fermions.
\begin{figure}[t]
\begin{center}
\includegraphics[width=7.6cm]{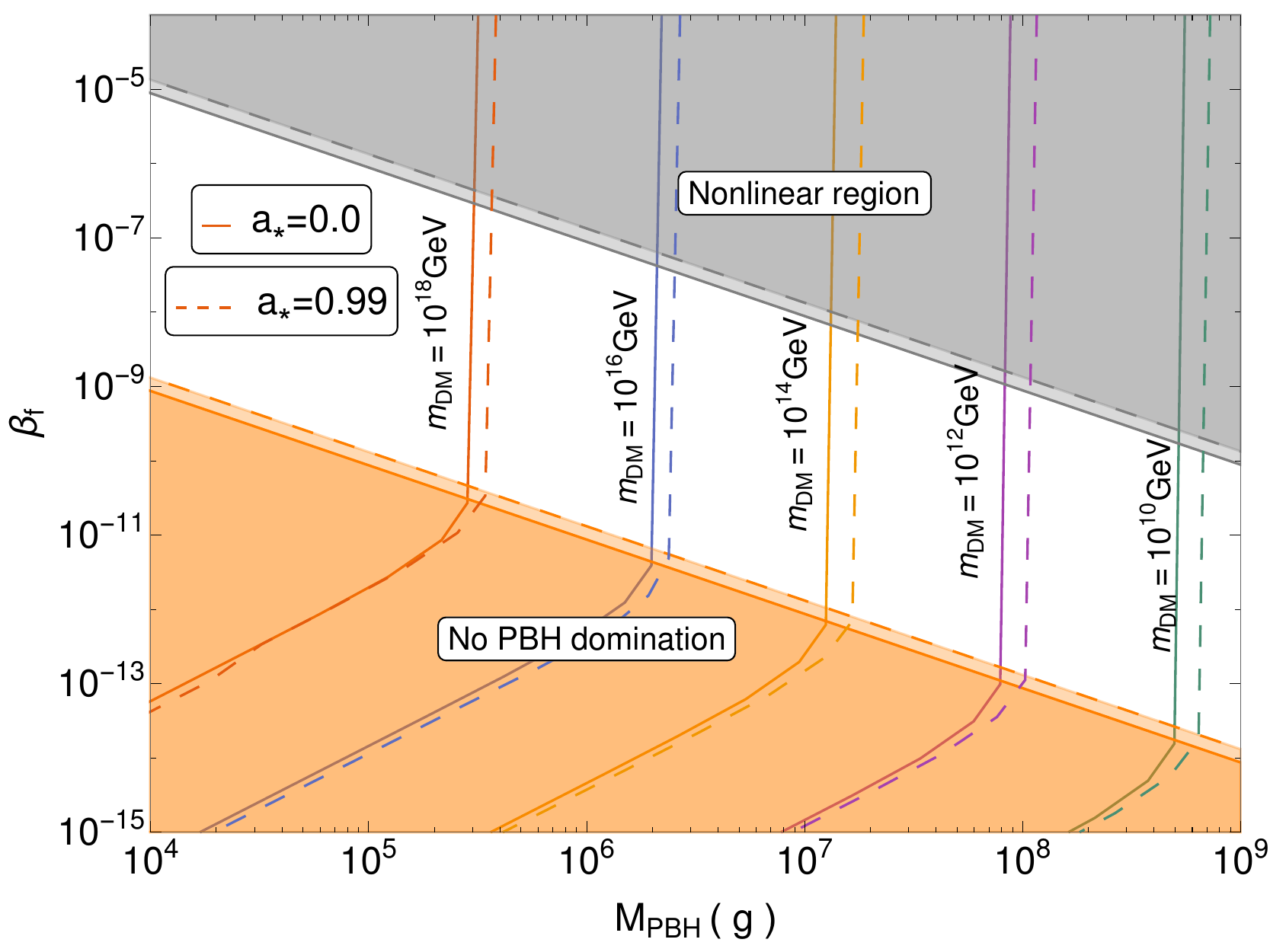}
\hskip 2pt
\includegraphics[width=7.6cm]{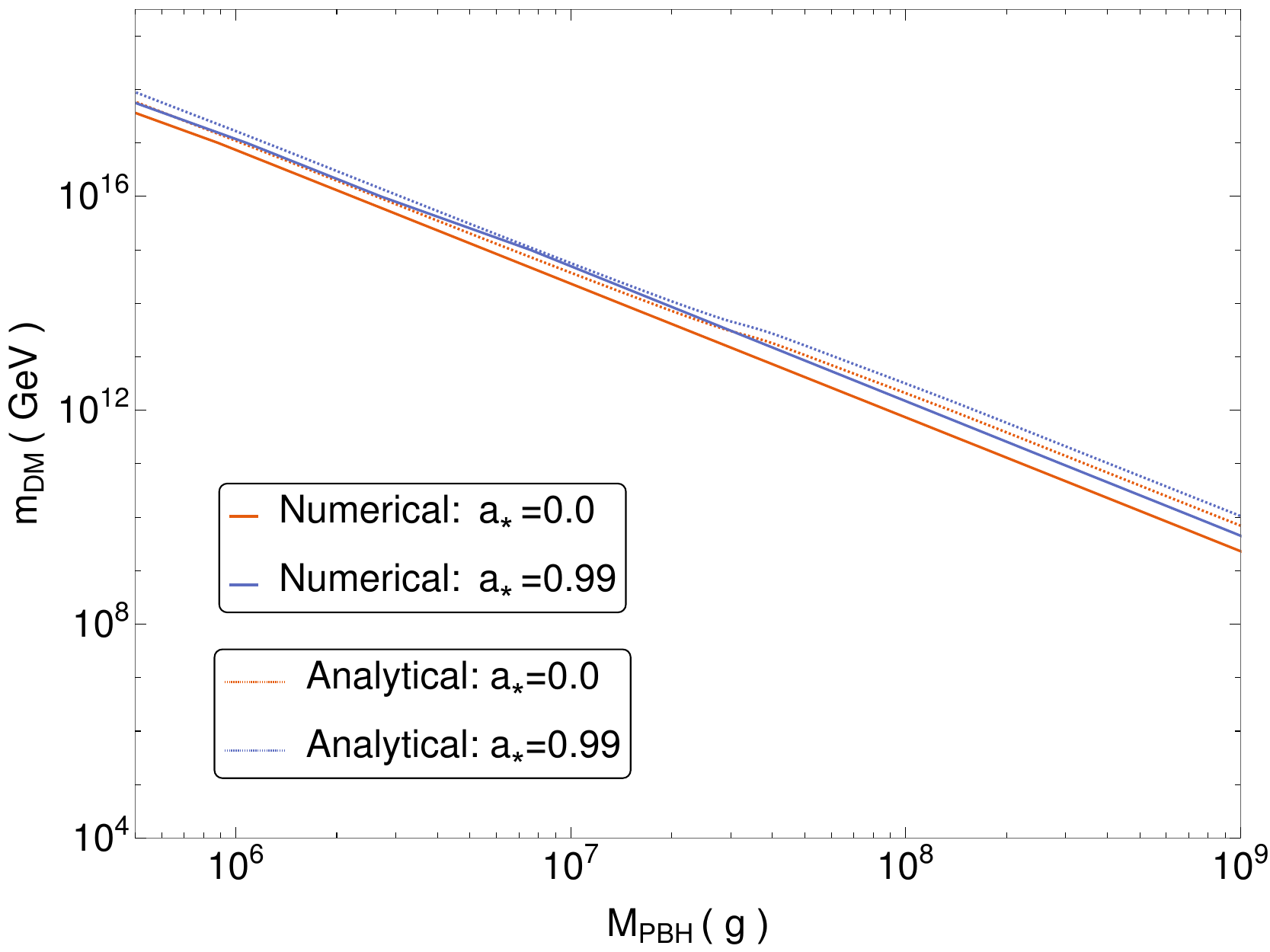}
\vskip 10pt
\caption{\textbf{Left panel:} We plot the values of the initial PBH abundance, $\beta_f$ as a function of PBH mass for the different masses of the DM particles, which contributes to the observed DM density of our universe ($\Omega_{DM} h^2 \approx 0.12$ \cite{Planck:2018vyg}). The non-linear regime ($\tau_{\rm rat} \ge 470$) is shaded in grey, and no PBH domination region is in orange.
\textbf{Right panel:} We set $\beta_f$ such that we get an era of PBH domination and plot the mass of the heavy DM particles, satisfying the observational DM relic density. We also find good agreement between the analytical result for $m_{\rm DM}$ (as in eqn. \ref{mdmanaly}) and exact numerical calculation.}
\label{con24}
\end{center}
\end{figure} 
Our approach to estimating the present-day DM relic density has been twofold. We use a numerical setup for exact estimates and compare it with independently derived approximate analytical results.

To take into account the continuous production of DM particles from Hawking evaporation and simultaneously the dilution effects from the expansion of the universe, we implement \texttt{FRISBHEE}, which solves Friedmann's equations to estimate the DM energy density at the start of standard RD epoch,  primarily contributed at the very last stage of PBH evaporation, when the universe goes through a transition from PBH domination to standard RD. The present-day DM energy density is then obtained using the ratio between then and present-day temperature and the change in the total number of relativistic degrees of freedom. From the conservation of entropy, we can write, 
\be 
a(t_1)^3 g_{*,S}(T_1) {T_1}^3 =  a(t_2)^3 g_{*,S}(T_2) {T_2}^3 \implies \left( \frac{a(t_1)}{a(t_2)} \right) = \left( \frac{T_2}{T_1} \right)\left( \frac{g_{*,S}(T_2)}{g_{*,S}(T_1)} \right)^{1/3} \, .
\ee
Using this, the DM energy density $\rho_{\rm DM}(t_{\rm r})$ of the start of RD ($t=t_r$) can be translated to present day ($t=t_0$) DM energy density $\rho_{\rm DM}(t_{\rm 0})$ as,
\be
\left( \frac{\rho_{\rm DM}(t_{\rm 0})}{\rho_{\rm DM}(t_{\rm r})} \right)=\left( \frac{a(t_r)}{a(t_0)} \right) = \left( \frac{T_0}{T_r}\right)^3 \left( \frac{g_{*}(T_{\rm 0})}{g_{*}(T_{\rm r})}\right).
\ee
\begin{figure}[!]
\begin{center}

\includegraphics[width=12cm,height=7.6cm]{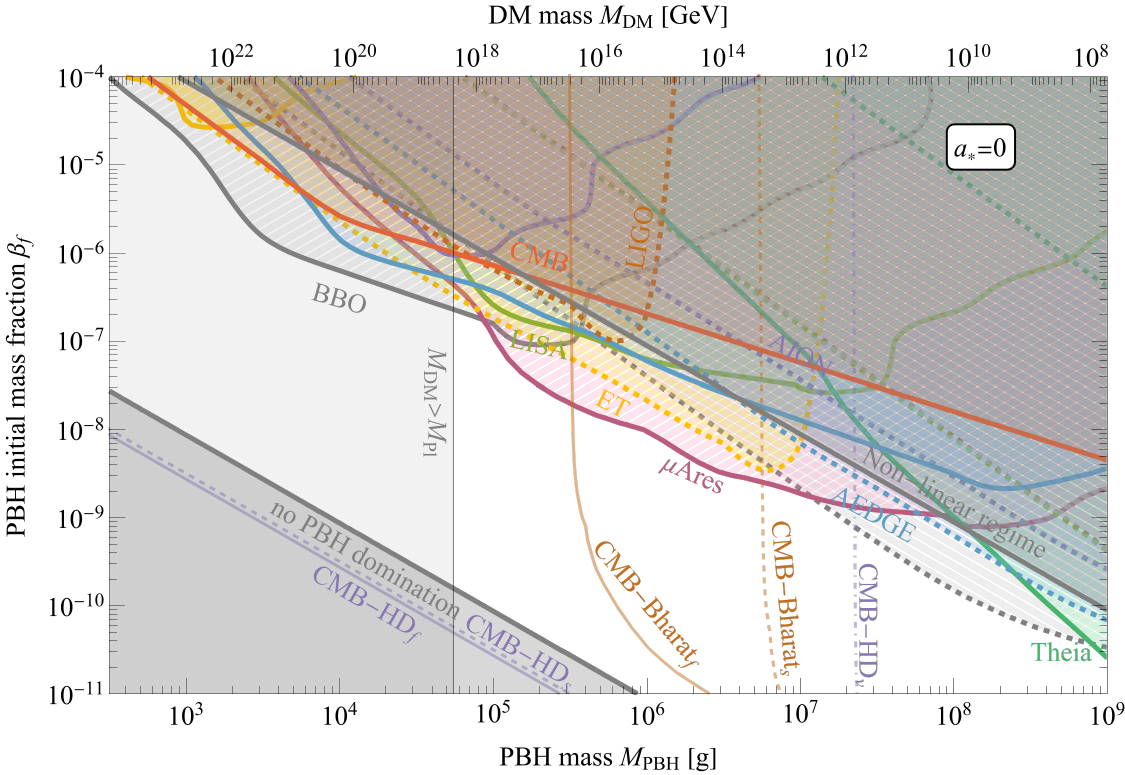}
\includegraphics[width=12cm,,height=7.6cm]{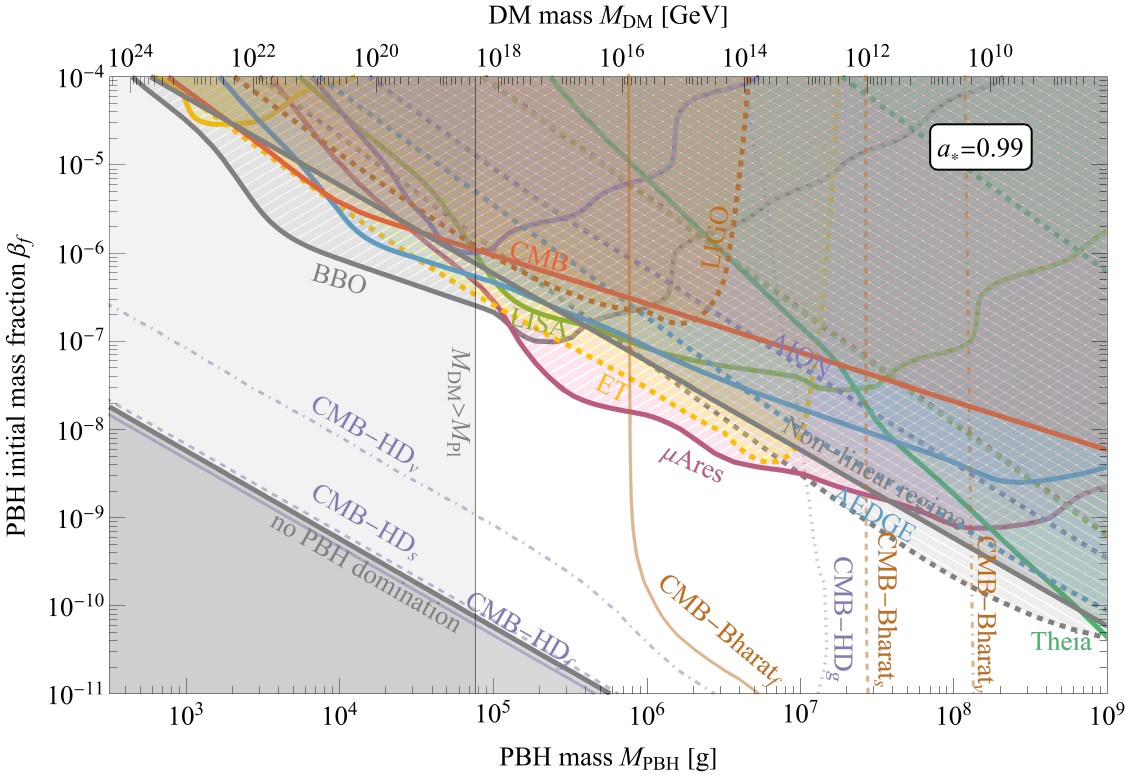}
\vskip 5pt
\caption{
Filled contours indicate the spectra within the detection range of future GW experiments similar to Fig.~\ref{spincontours}. 
%. The detection of each of our two peaks is indicated in a separate way for each experiment: low-frequency inflationary adiabatic contribution detection is shown by '$/$' dashed filling and solid contours, while the high-frequency isocurvature-induced adiabatic peak is indicated by '$\backslash$' dashed filling and dashed contours. 
%Contours without filling indicate parts of the parameter space in reach of the future CMB observatories due to the DR produced. 
The top ticks indicate the mass of the DM relic leading to the correct abundance today.
The top panel corresponds to $a_*=0.0$, while in the lower panel, $a_*=0.99$. Our DM results are only valid to the right of the vertical grey lines indicating $M_{\rm DM} = \Mpl$.
For contours denoted by CMB-HD or CMB-Bharat (without filling), the higher $\MPBH$ and higher $\beta_f$ region refers to the parameter-space where $\Delta N_{\rm eff}$ from DR is within reach of future CMB experiments.
 We show scalar, fermionic, vector, or graviton DR contours indicated by the subscripts $s$,$f$,$v$, or $g$, respectively, just as in  Fig.~\ref{spincontours}.
}
\label{con22}
\end{center}
\end{figure}
We numerically scan the PBH parameter space, namely the PBH mass and initial abundance considering the different mass of DM particles $m_{\rm DM}$ which can produce the observed DM abundance $\Omega_{\rm DM} h^2 \approx 0.12$ \cite{Planck:2018vyg}, for two different initial spin values $a_*= \left\{0.0, 0.99\right\}$, as shown in the left panel of Fig. \ref{con24}. 

We find that the change in the initial spin leaves a small but nonzero shift in the contours, which connects the points in PBH parameter space corresponding to a particular DM particle mass, contributing to the observed DM relic density. Another interesting point is the exact vertical part of these lines in the region where PBH domination occurs. This lack of dependence on the PBH initial abundance is expected because in the case of PBH domination, the total energy density just before the standard RD phase is contributed only from the PBHs, and the fraction of DM energy density remains constant for a fixed mass DM particle. In the case when PBHs evaporate before they can dominate, the fraction of PBH contribution to the total energy density decreases, which also leads to a decrement in the DM energy density coming from PBHs.

We depend upon a few reasonable simplifying assumptions for the analytical estimate of present-day DM relic density contributed by PBH evaporation. The energy density of the DM particles evolves similarly to PBHs during PBH domination. For spin 1/2 particles, we effectively assume all the DM particles are produced instantly at the end of the PBH evaporation process as their production rate tends to peak at the very end \cite{Masina:2021zpu}. If we consider the mass of the fermionic DM particle to be $m_{\rm DM} \gg T_{\rm PBH,i}$, the total number of DM particles from a single BH can be approximated as  \cite{Lennon:2017tqq},
\begin{equation}
N_{\rm DM} \approx 7.27\times 10^{33} g_{\rm DM} \left( \frac{\rm GeV}{m_{\rm DM}}\right)^2 \, .
\label{eq:particlenumber}
\end{equation}
At the end of PBH evaporation which roughly contributes to DM energy density,
\be
\rho_{\rm DM}(\tau_{\rm r})=\rho_{\rm tot}(\tau_{\rm r}) \left( \frac{N_{\rm DM} m_{\rm DM}}{\MPBH} \right)
\ee
Translating it to the present-day energy density, we get
\be
\rho_{\rm DM}(\tau_{\rm 0})=\rho_{\rm DM}(\tau_{\rm r}) \left( \frac{T_0}{T_r}\right)^3 \left( \frac{g_{*}(T_{\rm 0})}{g_{*}(T_{\rm r})}\right)
\ee
and the present DM relic density of our universe comes out to be,
\begin{eqnarray}
\Omega_{\rm DM} =\frac{\rho_{\rm DM}}{\rho_{\rm crit}} \approx \bigg(\frac{g_{{\rm DM}}}{g_{*,H}(T_r)}\bigg) \, \bigg(\frac{3.6 \times 10^{9} \, {\rm GeV}}{m_{\rm DM}}\bigg) \, \bigg(\frac{10^8 \, {\rm g}}{M_{\rm PBH}}\bigg)^{5/2} (g_{*}(T_r))^{3/4} \, ,
\end{eqnarray}
where we use the temperature of the universe at the end of PBH evaporation,
\be 
T_r \simeq 8.9\times 10^{7} 
\left(\frac{\MPBH}{10^2 g} \right)^{-3/2} ({g_*(T_r)})^{-1/4}~ {\rm GeV} \, .
\ee
While this estimate is valid for non-spinning PBHs, it is tough to estimate $N_{\rm DM}$ analytically for spinning PBHs. From our results in section \ref{darkradiation} that the emission of fermions stays nearly unaffected due to the change in initial PBH spin, we use this as an assumption to get an analytical estimation of $\Omega_{\rm DM}$ for spinning PBHs. Taking $N_{\rm DM, spin} \simeq N_{\rm DM, no\text{-}spin}$, allows us to express,
\be
\rho_{\rm DM, spin}(\tau_{\rm 0}) \simeq  \rho_{\rm DM, no\text{-}spin}(\tau_{\rm 0}) \left( \frac{T_{\rm r, spin}}{T_{\rm r, no\text{-}spin}}\right)\, ,
\ee
and
\be
\left( \frac{\Omega_{\rm DM, spin}}{\Omega_{\rm DM, no\text{-}spin}} \right) \approx  \left( \frac{T_{\rm r, spin}}{T_{\rm r, no\text{-}spin}}\right) \approx \left( \frac{\Delta t_{\rm PBH, spin}}{\Delta t^S_{\rm PBH}}\right)^{-1/2} = ~{\mathcal{F}(a_*)}^{-1/2}.
\ee

It is also possible to get a significant abundance of DM particles even if we consider the scenario wherein the PBHs evaporate before they can dominate, as we can see in the lower part of the left panel of Fig. \ref{con24}.
Since our focus here is to connect DM with the observable ISGWB signals and we obtain amplification in the ISGWB spectrum only when there is a phase of PBH domination, for the analytical estimates, we limit ourselves to the parameter space where PBH can dominate the universe for a finite duration. Matching with the observed abundance of DM relic density provides us with a relation between the mass of PBHs and DM particle mass,
\be 
m_{\rm DM} \approx 1.7 \times 10^{5} \left( \frac{10^{10}\, \text{g}}{\MPBH} \right)^{5/2} ~{\mathcal{F}(a_*)}^{-1/2}~ \text{GeV}.
\label{mdmanaly}
\ee
Using this expression, we plot $m_{\rm DM}$ as  a function of $\MPBH$ in the right panel of Fig. \ref{con24} along with the corresponding numerical values, which are obtained from scanning the parameter space where PBHs do dominate the universe for a finite period. The numerical and analytical results match quite closely, and we use the analytically derived $m_{\rm DM}$ in the upper bars of the upper and lower panel of Fig. \ref{con22} for $a_*=\left\{0.0, 0.99\right\}$. To complete the picture, we plot it along with the contours for detectable GW observations (with SNR $\geq 10$) and the CMB observation bounds for $\Delta N _{\rm eff}$ for different particles species with $s=0,1/2,1,2$ as we have derived in the previous section.

\subsection{Baryogenesis}
\begin{figure}[t]
\begin{center}
\includegraphics[width=7.4cm]{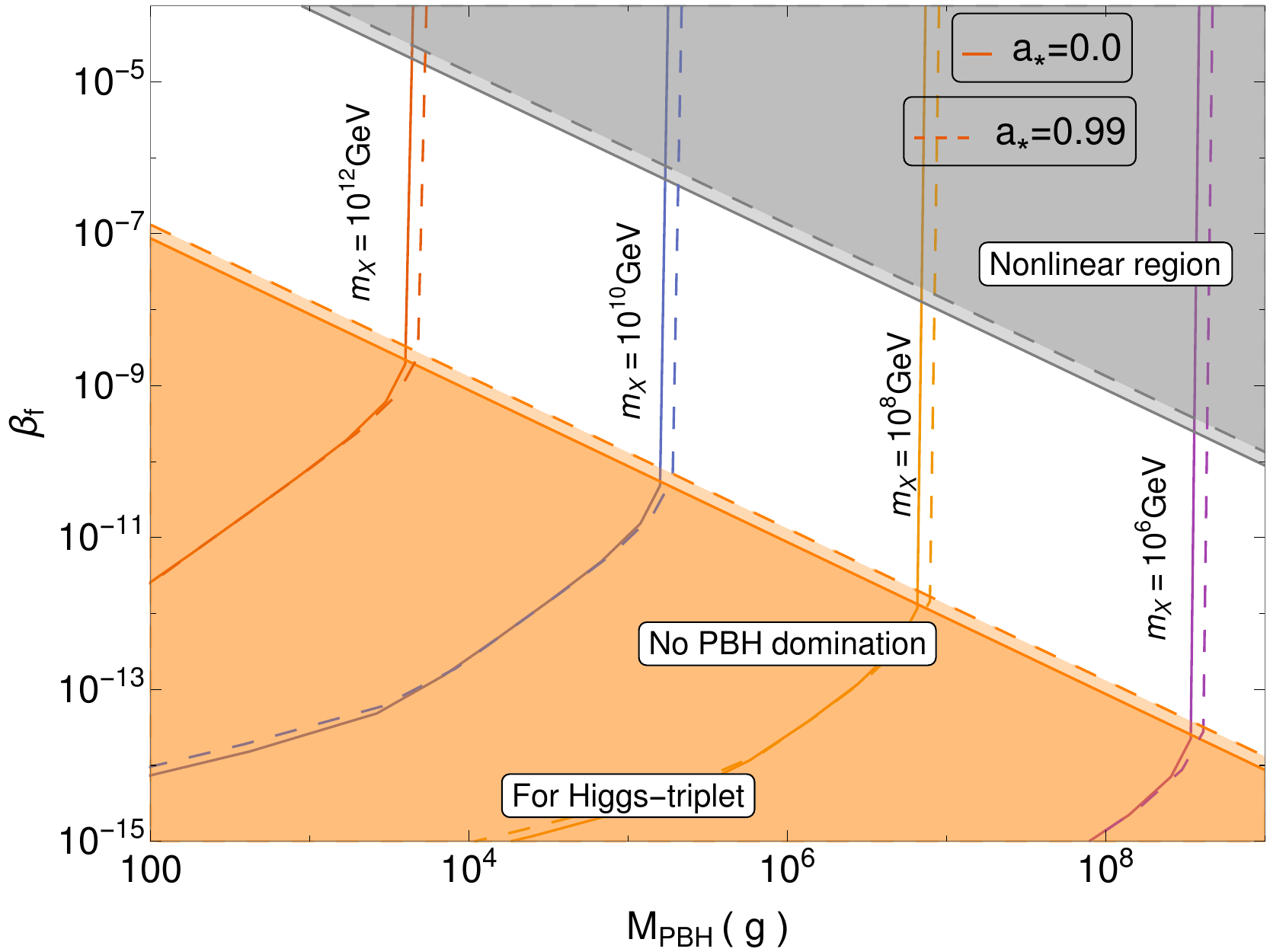}
\hskip 2pt
\includegraphics[width=7.4cm]{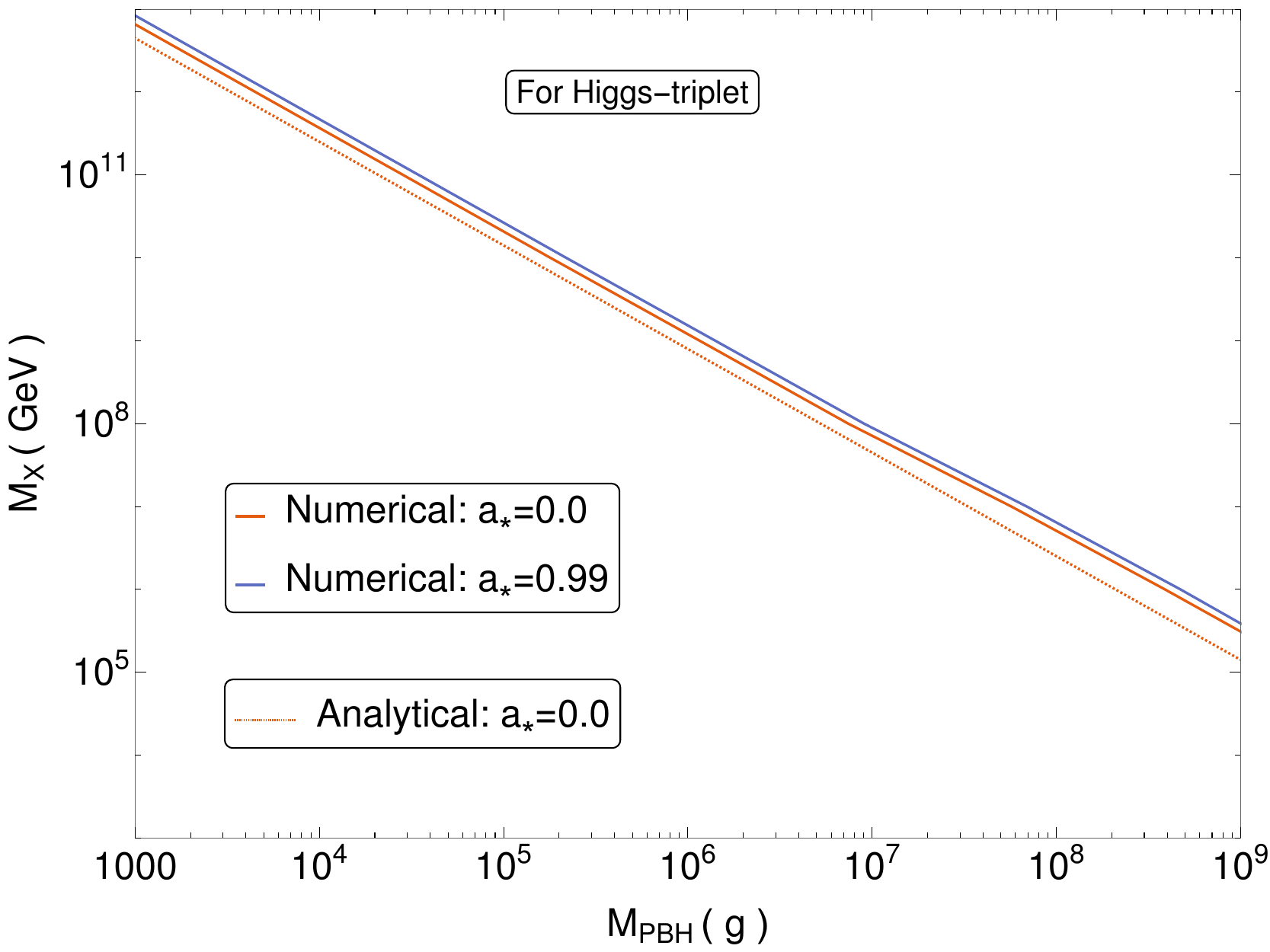}
\hskip 2pt
\includegraphics[width=7.4cm]{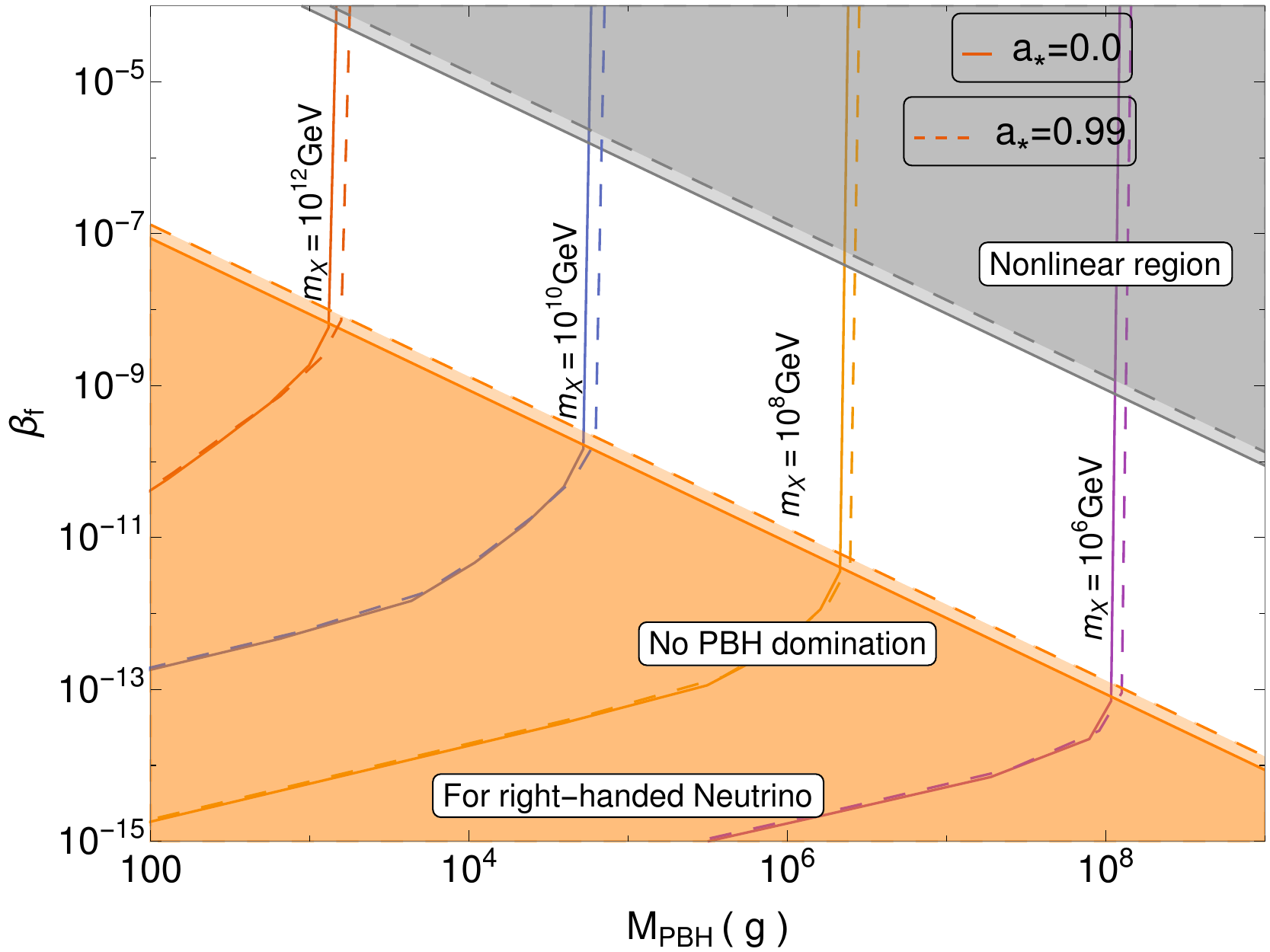}
\hskip 2pt
\includegraphics[width=7.4cm]{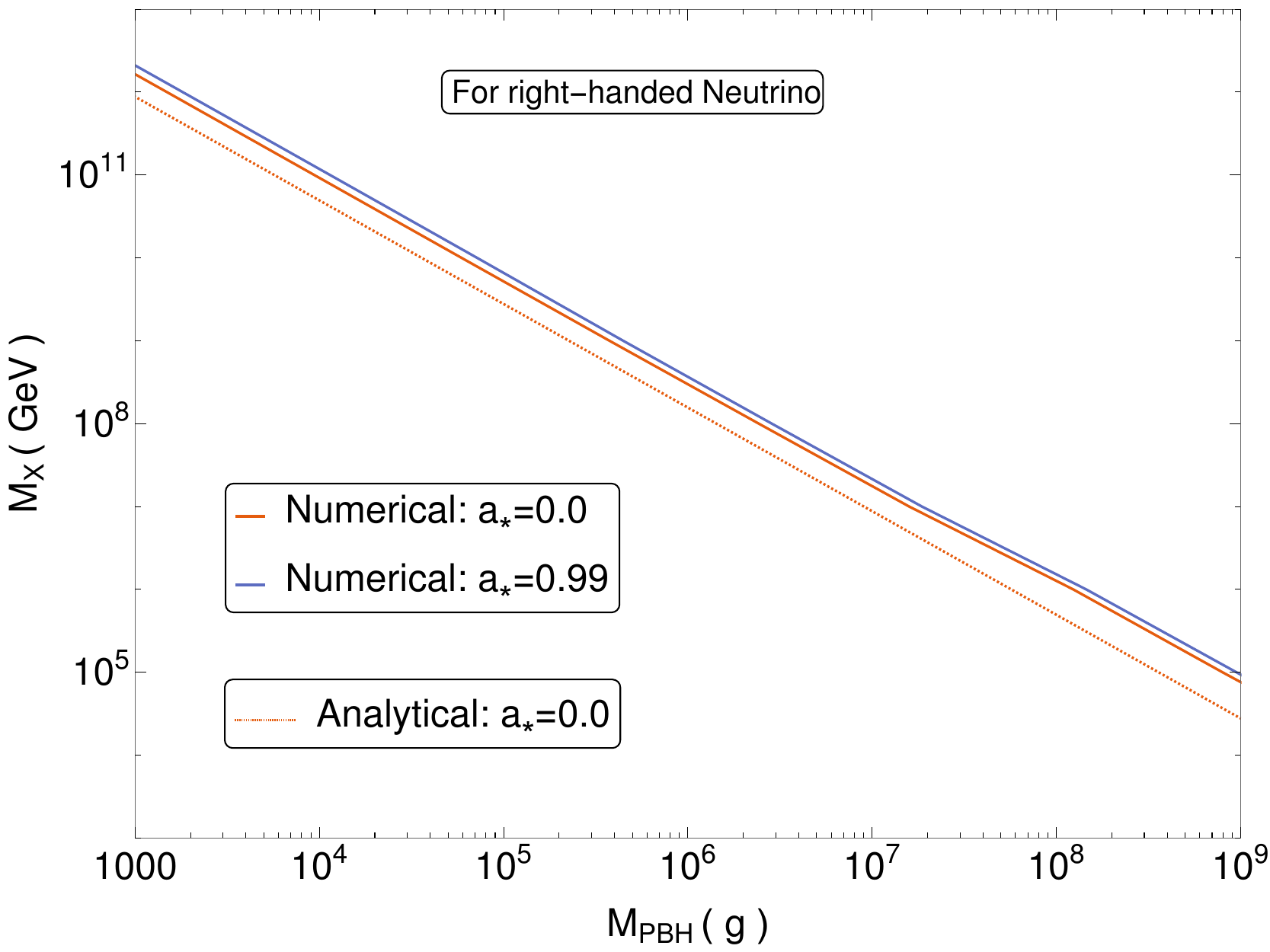}
\vskip 10pt
\caption{\textbf{Left panel:} We plot $\beta_f$ as a function of $\MPBH$ for the different masses of the heavy decaying particles, which contributes to the observed baryon asymmetry of our universe ($Y_{B} \approx 8.80\times 10^{-11}$). The non-linear regime ($\tau_{\rm rat} \ge 470$) is shaded in grey and no PBH domination in orange.
\textbf{Right panel :} We set $\beta_f$ such that we get an era of PBH domination and plot the mass of the heavy decaying particles which satisfy the baryon asymmetry observation. The comparison between the analytical and numerical results for non-spinning PBHs is reasonably good.}
\label{con23}
\end{center}
\end{figure}
\begin{figure}[t]
\begin{center}
\includegraphics[width=12cm]{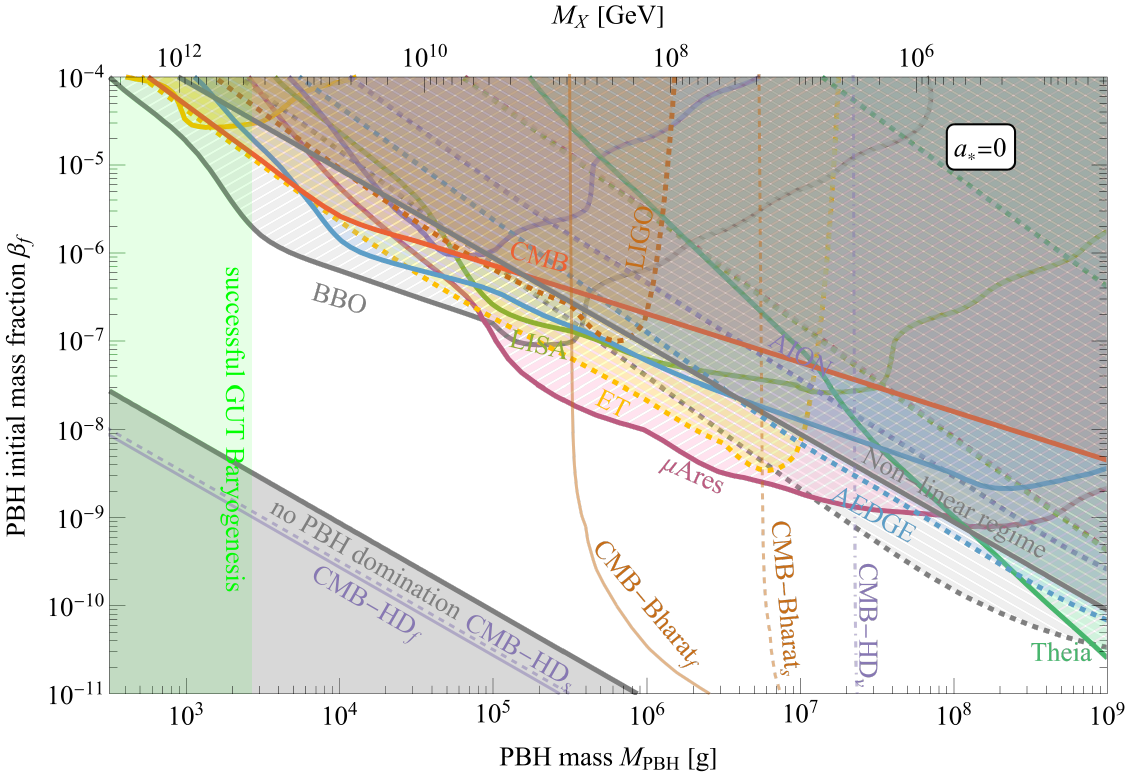}
\vskip 6pt
\includegraphics[width=12cm]{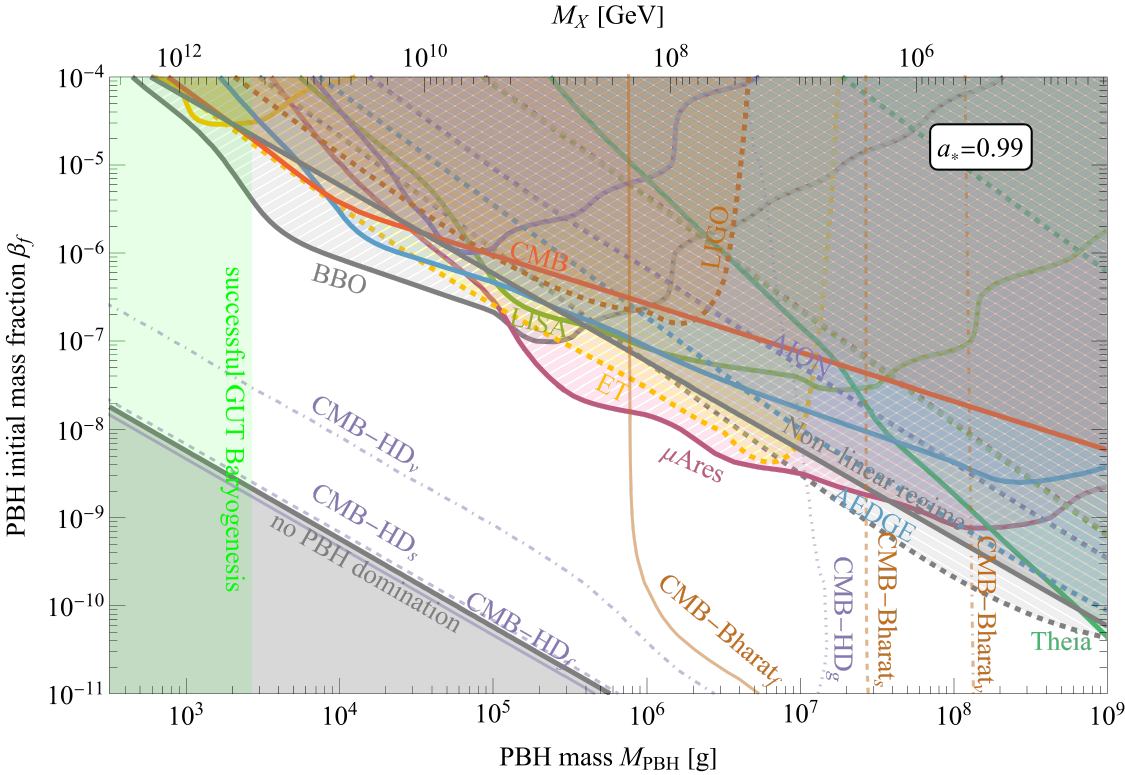}
\vskip 10pt
\caption{Reach of future GW observatories (filled contours) and CMB probes for DR (contours denoted by CMB-HD or CMB-Bharat, without filling) in terms of the mass and initial abundance of PBHs, just as in Fig.~\ref{con22}. 
%Also contours denoted by CMB-HD or CMB-Bharat (without filling), refer to the parameter space where $\Delta N_{\rm eff}$ from DR is within reach of future CMB experiments similar to Fig.~\ref{con22}.
The top panel corresponds to $a_*=0.0$ while the lower is for $a_*=0.99$. On the top axis, we indicate the mass of the Higgs triplet $m_X$ for which we obtain the observed baryon asymmetry. The green shaded region corresponds to allowed regions for GUT baryogenesis from proton decay bound. }
\label{con2}
\end{center}
\end{figure}

As we discussed in the previous section, massive non-relativistic stable particles emitted from Hawking evaporation of PBHs can lead to the DM component of our universe. On the other hand, if the emitted massive particles are unstable, they can decay in a baryon-number violating process which can contribute to the observed baryon asymmetry of our universe.

In our previous work \cite{Bhaumik:2022pil}, we considered PBH evaporation induced baryogenesis and connected the baryogenesis parameters with the PBH parameter space in the light of ISGWB observations. We considered two cases of baryogenesis: direct baryogenesis through the baryon-number violating decay of Higgs-triplets
and  baryogenesis through leptogenesis due to the decay of right-handed neutrinos and the subsequent conversion of lepton asymmetry to baryon asymmetry via the sphaleron processes.
We assumed the baryogenesis scenario, where Higgs-triplet 
and right-handed neutrinos
originate from Hawking evaporation of ultra-low mass PBHs. We denote the efficiency of the baryogenesis process with a parameter $\epsilon_X$, which can take different values up to unity depending on various decay channels and other factors in the baryogenesis process solely depending on the microphysics of the BSM particle physics scenario under consideration. One can write an expression for $\epsilon_X$ in terms of the decay rate of particle model \cite{Hooper:2020otu},
\begin{equation}\label{epsilon}
\epsilon_X \equiv \sum_i B_i \frac{\Gamma (X\to f_i) - \Gamma(\bar{X} \to 
\bar{f}_i)}{\Gamma_{\rm tot}}\, ,
\end{equation}
where $f_i$ is $i^{th}$ final particle with baryon number $B_i$, and $\Gamma_{\rm tot}$ is the total decay width of the heavy particle. For the baryogenesis via leptogenesis scenario, we denote the  conversion factor for lepton asymmetry to baryon asymmetry as $\lambda = 0.35$ \cite{Fujita:2014hha}.

In the present work, we revisit the same framework to include the effects of the initial spinning PBHs. For this purpose,  we modify \texttt{FRISBHEE} to calculate the number density of the Higgs-triplet particle, $N_{X}(t_r)$ numerically at the end of PBH evaporation and estimate the total baryon asymmetry from Higgs-triplet particles as,
\be
\label{BHdom0}
Y_B = \frac{n_{\rm PBH}}{s(t_r)} \epsilon_{X} N_{X}=\left( \frac{N_{X}}{\MPBH} \right) \frac{\rho_{\rm tot}(t_{\rm r}) }{s(t_r)} \epsilon_{X}= \left( \frac{3}{4} \right) \left( \frac{N_{X}T_{\rm r}}{\MPBH} \right) \frac{g_{*}(T_{\rm r}) }{g_{*,S}(T_{\rm r}) } \epsilon_{X}\, .
\ee
For Baryogenesis via leptogenesis mechanism with right-handed neutrinos, we need to include the conversion factor $\lambda$, and this formula gets modified to,
\be
\label{BHdom}
Y_B = \lambda \left( \frac{3}{4} \right) \left( \frac{N_{X}T_{\rm r}}{\MPBH} \right) \frac{g_{*}(T_{\rm r}) }{g_{*,S}(T_{\rm r}) } \epsilon_{X}\, .
\ee
In the left panels of Fig. \ref{con23}, we use the exact numerical setup and plot the lines connecting the points of PBH parameter space which are found to produce the observed baryogenesis for a particular mass of decaying particles, Higgs-triplet (upper panel) and right-handed neutrinos (lower panel). For the case of PBH domination, we find the estimated baryon asymmetry to have negligible dependence on initial PBH abundance $\beta_f$, which is qualitatively similar to the DM and DR scenario.  Using $Y_B  \approx 8.8 \times 10^{-11} $  (inferred from CMB observations 
 \cite{Planck:2015fie, Planck:2018vyg} and BBN constraints \cite{ Alvey:2019ctk, ParticleDataGroup:2018ovx} )
and $\epsilon_X=1$, we get an analytical relation between the initial PBH mass $M_{\rm PBH}$ and mass of the decaying heavy particle $M_{X}$ as \cite{Bhaumik:2022pil},
\be
\label{MX1}
M_{X} \approx 2.5 \times 10^{16} \sqrt{\epsilon_X\left(\frac{1 ~\text{g}}{M_{\rm PBH}}\right)^{5/2}} ~\text{GeV} \hspace{.5cm} \text{(For Higgs-triplet in GUT baryogenesis)}.
\ee
\be
\label{MX2}
M_{X} \approx 5.9 \times 10^{15} \sqrt{\epsilon_X\left(\frac{1 ~\text{g}}{M_{\rm PBH}}\right)^{5/2}} ~\text{GeV} \hspace{01.9cm} \text{(For right-handed neutrinos )}.
\ee
In the right panels of Fig. \ref{con23}, we plot these approximate analytical results for the Schwarzschild case, along with the numerically obtained mass of the Higgs-triplet (upper panel) and right-handed neutrinos (lower panel) as a function of $\MPBH$ which leads to observed matter-antimatter asymmetry, both for initially spinning and non-spinning population of PBHs. We find a negligible deviation from non-spinning results to spinning PBHs, while the mismatch between numerical and analytical results is somewhat significant.

As we are considering scalar Higgs-triplet particles in the GUT baryogenesis context and right-handed neutrinos in case of baryogenesis via leptogenesis scenario and the emission of $s=0$ and $s=1/2$ particles get negligibly affected due to the inclusion of spin, a very weak dependence on the initial PBH spin is expected. Using our analytical Schwarzschild expression, we plot the corresponding mass ranges $m_X$ of Higgs triplet particles in the upper panel of Fig. \ref{con2}. We also plot the proton decay bound in the green vertical contours for direct GUT baryogenesis models as required by the laboratory constraints on proton decay from SuperK experiment~\cite{Super-Kamiokande:2016exg} and take the efficiency parameter $\epsilon_X=1$ for all our calculations. The lower panel of Fig. \ref{con2} show similar contours for spinning PBHs with $a_*=0.99$. It is interesting to note the comparative difference in the ISGWB detection possibility lines in the two plots for two different values of the spin parameter $a_*$. However, the predictions from baryogenesis remain nearly the same. It is important to note that we only plot the mass range of Higgs-triplet particles for direct baryogenesis in the upper panel of Fig. \ref{con2}, which is not valid for right-handed neutrinos. We plot it as a representative plot to illustrate the combined picture of ISGWB detection probability, detectable DR parameter space, and baryogenesis scenario.
%%%%%%%%%%%%%%%%%%%%%%%%%%%%%%%%%%%%%%%%%%%%%%%%%%%%%%%%

\section{Conclusions and discussions  }
\label{discuss}

In this paper, we have explored diverse observational consequences of an early scenario populated by ultra-low mass PBHs, which briefly dominate the energy density of the universe but evaporate before BBN. 
There are several important implications of such a scenario; for instance, the emitted byproducts from PBHs, depending on their properties, can contribute to the DM energy density, additional DR contribution and baryogenesis via the decay of unstable particles.
In this work, we study how these different aspects can complement each other, and a combination of them, as a whole, can offer a better understanding of the post-inflationary phase before BBN. In our earlier work \cite{Bhaumik:2022pil},
we studied the origin of a doubly-peaked ISGWB in such a scenario which can be used as a novel  tool to probe such a possibility. Here, we have extended the scope of our earlier work by including the initially spinning PBHs and quantitatively identified the corresponding signatures in DR, DM, and baryogenesis with the help of an exact numerical setup. Some interesting findings of our work can be summarized below:

\begin{itemize}
    \item
    %\textbf{Effects of spin on ISGWB :} 
 The resonant ISGWB spectrum of a unique doubly-peaked shape contributed from the isocurvature-induced adiabatic perturbations associated with the PBHs distribution and from the inflationary adiabatic scalar perturbations can be detected with various future GWs detectors (see, Figs. \ref{spincontours} and \ref{con22}) and can also be used to constrain the PBH parameter space from the results on $N_{\rm eff}$ from the existing CMB observations. For the ISGWB, the primary effect of including the initially spinning PBHs arises from the modified PBH lifetime. PBHs with non-zero spin evaporate more rapidly, and that change in the evaporation timescale results in the modification of the ISGWB spectrum as shown in Figs.~\ref{sps-gw2} and \ref{sps-gw0}. The peak frequency and the height of the peak both get modified slightly, which shows a weak dependence of ISGWB on the PBH spin.

    \item 
   % \textbf{Spin effects for dark radiation and ISGWB : } 
    While the PBH spin does not play a significant role in the ISGWB signal, in the case of the DR from Hawking evaporation, the effect of initially spinning PBHs varies depending on the type of emitted particle, as we can see from Figs. \ref{DRR1} and \ref{DRR2}, from scalar particles, larger PBH spin results in smaller $\Delta N_{\rm eff}$ and reduced detectability, but as we go for higher spin DR particles like the graviton, the larger PBH spin is found to contribute significantly to $\Delta N_{\rm eff}$, thereby, enhancing the detectability.
    This characteristic feature of gravitons opens up an exciting possibility of detecting the initial PBH spin through DR in future CMB observations like CMB-HD and CMB-Bharat. As evident from Fig. \ref{spincontours}, we obtain a PBH mass-dependent cutoff value of PBH spin, only above which $\Delta N_{\rm eff}$ from the graviton becomes detectable in the CMB-HD experiment. Though the sensitivity of PBH spin is weaker in the case of vector particles compared to the graviton, similar behaviour for the $\Delta N_{\rm eff}$ detectable in CMB-HD is also visible for vector particles.

    \item 
    %\textbf{Spin effects for dark matter and baryogenesis with ISGWB : }
    Non-thermal production of DM candidates satisfying the observed relic density is notoriously challenging to test in DM laboratory experiments due to the null interactions with the standard model as well as the high energy scales involved. However, the doubly-peaked ISGWB spectrum allows us to probe such DM production with varying DM masses, as shown in Figs.~\ref{con24} and \ref{con22}. Considering Dirac fermions as DM particles, we find the dependence of PBH spin to be very weak. Similarly, for PBH-induced baryogenesis, we find a very weak PBH spin sensitivity both for direct baryogenesis and baryogenesis via leptogenesis scenario, as evident from Fig. \ref{con23}.

    \item 
    %\textbf{ CMB complementarity :} 
    Though initial PBH spin leaves insignificant effects in the case of ISGWB, fermionic DM, and scalar vector and fermionic DR particles, we find significant effects of PBH spin for graviton as DR particles. It offers us an exciting opportunity to employ otherwise unrelated experiments, CMB detectors, and GW detectors to break the degeneracy between spinning and nonspinning PBH scenarios. Thus, cosmological estimations of $\Delta N_{\rm eff}$ from CMB experiments and detection of ISGWB in future GW observations complement each other (see Fig. \ref{con24}) to constrain the reheating history more effectively.

    \item 
    %\textbf{Effectiveness of different probes to measure and constrain the initial PBH abundance and the Non-linear region: }
    Different observational probes offer very different sensitivities for initial PBH abundance, $\beta_f$. As we have discussed, ISGWB strongly depends on the initial PBH abundance, as even a slight change in $\beta_f$ alters the duration of the early PBH domination and leads to large modifications of the ISGWB peaks. On the other hand, when we consider scenarios with the generation of DR, DM or baryogenesis from Hawking evaporation, we find them to depend on $\beta_f$ only when PBHs evaporate before they can dominate the universe.
    But, if the PBHs dominate the universe for a finite duration, the initial abundance of PBH or the duration of PBH domination does not alter the results significantly, as is evident from the left panels of Figs. \ref{con24} and \ref{con23}, and also from the sensitivity plots in Figs. \ref{con22} and \ref{con2}.

\end{itemize}

To conclude,  Hawking evaporation of a dominant population of ultra-low mass spinning and non-spinning PBHs can lead to cosmological relics such as DM, DR, and baryon asymmetry, and the doubly-peaked ISGWB arising inevitably in such a scenario can potentially probe and constrain these contributions. We have  exhibited a novel complementarity of different probes like the future CMB experiments and GW detectors and found the synergy of future GW observatories and CMB missions to observe $\Delta N_{\rm eff}$ to put very strong constraints on PBH parameter space as well as the reheating history of the pre-BBN universe.

There have been some recent efforts to include more realistic non-monochromatic PBH mass distribution, both in the context of ISGWB \cite{Papanikolaou:2022chm} and on the Hawking evaporation-induced baryogenesis \cite{Gehrman:2022imk}. In a similar context, it would also be interesting to explore the effects  of the non-monochromatic spin distribution of ultra-low mass PBHs and resulting modifications in ISGWB, dark sector and baryogenesis. Taking into account the clustering of PBHs during the eMD epoch and different spatial distributions of PBHs are also expected to leave imprints on our results, particularly on the ISGWB part, which we leave for future work. 
We do not work with any PBH-forming models for the inflationary adiabatic perturbations. Instead, we use the standard power-law power spectrum on all scales. If we consider specific inflationary models for the formation of such ultra-low mass PBHs, our estimation of the ISGWB near the PBH forming scales is expected to be modified, which would be interesting to explore further.
A prolonged eMD phase also restricts the allowed number of e-folds from the horizon exit of the CMB pivot scale to the end of inflation \cite{Bhaumik:2020dor}. This constraint, along with the optimal range of the scalar spectral index $n_s$ and the tensor to scalar ratio $r$ from CMB observations \cite{Planck:2018jri} allows one to put constraints on possible inflationary models.
%\ag{and baryogenesis scenarios in this regard \cite{Ghoshal:2022fud}.} \nb{no, i think the constrain is primarily on PBH parameter space, then you can translate it to all other ISGWB DR DM ..etc i think the paper you are citing depends on the reheating model rather than the constrain on the efold number, which is quite different from what i want to say }. 
%In the context of general reheating setup, similar considerations of e-fold numbers can constrain the different models of leptogenesis \cite{Ghoshal:2022fud}. 
For an early PBH domination, such an analysis would allow us to constrain the PBH parameter space directly and, thereby, check the consistency of the Hawking evaporation-driven baryogenesis and DM production, for a particular model of inflation. We leave it for future work.

%This establishes the  future GW observatories as complemenatrity machines to future CMB missions to observe $\Delta N_{\rm eff}$ to test unexplored PBH-domination pre-BBN era in early universe, concerning PBHs with masses $\lesssim 10^9$~g.
%Our results are general and can be readily applied to a wide class of models.

%\ag{We need to cite figures and equations in the ....}
%Nilanjan : DR+ details of the code + Baryogenesis
%Anish : DM
%Conclusions and discussions
%%%%%%%%%%%%%%%%%%%%%%%%%%%%%%%%%%%%%%%%%%%%%%%%%%%%%%%%

%\ag{Need ?}
%\begin{enumerate}
%    \item Find out a way to incorporate the discussion of D Neff combing Fig. 2 and 5.
%\end{enumerate}

\section*{Acknowledgments}
The authors thank Lucien Heurtier for useful discussions and suggestions about  {\texttt{FRISBHEE}}. NB thanks Yashi Tiwari, Ranjan Laha, and Akash Kumar Saha for helpful discussions and suggestions. This work was supported by the Polish National Agency for Academic Exchange within Polish Returns Programme under agreement PPN/PPO/2020/1/00013/U/00001 and the Polish National Science Center grant 2018/31/D/ST2/02048. RKJ acknowledges  financial support from the new faculty seed start-up  grant of the Indian Institute of Science, Bengaluru; Science and Engineering  Research Board, Department of Science and Technology, Govt. of India, through the Core Research Grant~CRG/2018/002200 and the Infosys Foundation, Bengaluru, India, through the Infosys Young Investigator Award.

%%%%%%%%%%%%%%%%%%%%%%%%%%%%%%%%%%%%%%%%%%%%%%%%%%%%%%%%

%\input{bib}

%\newpage
\bibliographystyle{JHEP}
\bibliography{ref}
\end{document}